\documentclass [11pt]{article}
\usepackage{amsfonts}
\usepackage{graphicx}
\usepackage{amsmath,amssymb} 
\usepackage{latexsym}
\usepackage{mathrsfs}
\usepackage{bbold}
\usepackage{color}
\usepackage{slashed}
\definecolor{red}{rgb}{1,0,0}
\usepackage{cancel}
\usepackage{comment}
\usepackage{cite}
\usepackage{xcolor}

\definecolor{red}{rgb}{1,0,0}

\makeatletter
\def\section{\@startsection {section}{1}{\z@}{-3.5ex plus -1ex minus
 -.2ex}{2.3ex plus .2ex}{\large\bf}}
\def\subsection{\@startsection{subsection}{2}{\z@}{-3.25ex plus -1ex
minus -.2ex}{1.5ex plus .2ex}{\normalsize\bf}}
\makeatother
\makeatletter

\@addtoreset{equation}{section}

\makeatother

\def\bea{\begin{eqnarray}} \def\eea{\end{eqnarray}}
\def\be{\begin{equation}} \def\ee{\end{equation}} \def\nn{\nonumber}

\newcommand{\promille}{%
  \relax\ifmmode\promillezeichen
        \else\leavevmode\(\mathsurround=0pt\promillezeichen\)\fi}
\newcommand{\promillezeichen}{%
  \kern-.05em%
  \raise.5ex\hbox{\the\scriptfont0 0}%
  \kern-.15em/\kern-.15em%
  \lower.25ex\hbox{\the\scriptfont0 00}}

\usepackage[colorlinks,linkcolor=black,citecolor=blue,urlcolor=blue,linktocpage]{hyperref}

\begin{document}

\thispagestyle{empty}

\begin{center}

\vspace*{-.6cm}

\hfill SISSA 65/2014/FISI \\

\begin{center}

\vspace*{1.1cm}

{\Large\bf  General Three-Point Functions in 4D CFT}
\end{center}

\vspace{0.8cm}

{\bf Emtinan Elkhidir$^{a}$, Denis Karateev$^{a}$, and Marco Serone$^{a,b}$}\\

\vspace{1.cm}

${}^a\!\!$
{\em SISSA and INFN, Via Bonomea 265, I-34136 Trieste, Italy} 

\vspace{.1cm}

${}^b\!\!$
{\em ICTP, Strada Costiera 11, I-34151 Trieste, Italy}

\end{center}

\vspace{1cm}

\centerline{\bf Abstract}
\vspace{2 mm}
\begin{quote}

We classify and compute, by means of the six-dimensional embedding formalism in twistor space, all possible three-point functions in four dimensional conformal field theories
involving bosonic or fermionic operators in irreducible representations of the Lorentz group.
We show how to impose in this formalism constraints due to conservation of bosonic or fermionic currents.
The number of independent tensor structures appearing in any three-point function is obtained by a simple counting.
Using the Operator Product Expansion (OPE), we can then determine the number of structures appearing in 4-point functions with arbitrary operators.
This procedure is independent of the way we take the OPE between pairs of  operators, namely it is consistent with crossing symmetry, as it should be.
An analytic formula for the number of tensor structures for three-point correlators with two symmetric and an arbitrary  bosonic  (non-conserved) operators is found, which 
in turn allows to analytically determine the number of structures in 4-point functions of  symmetric traceless tensors.

\end{quote}


\newpage

\tableofcontents

\section{Introduction}

Conformal Field Theories (CFTs) play a fundamental role in theoretical physics. For instance, they are the starting and ending points of renormalization group flows in quantum field theories,
they describe second-order phase transitions in critical phenomena  and, by means of the AdS/CFT correspondence, they can help us in shedding light on various aspects of  quantum gravity and string theory.  
Thanks to the tight constraints imposed by the conformal symmetry,  CFTs are also among the few examples (if not the only one) of interacting quantum field theories 
where exact results are available without supersymmetry in any number of space-time dimensions. In particular, it is well-known that three-point functions of scalar primary operators are univocally determined by the conformal symmetry, up to a coefficient.  Three-point functions of arbitrary fields, not only scalars, are also fixed by conformal symmetry up to some coefficients. 
Most of the attention has been devoted to correlators involving traceless symmetric 
conserved operators, see e.g. refs.\cite{Osborn:1993cr,Erdmenger:1996yc,Maldacena:2011nz,Giombi:2011rz,Maldacena:2011jn,Zhiboedov:2012bm,Stanev:2012nq,Dymarsky:2013wla}. 
 More general correlators involving again traceless symmetric (conserved or not) tensors have recently been computed in ref.~\cite{Costa:2011mg}, while some other specific correlator with fermions was considered, e.g., in ref.\cite{Weinberg:2010fx}. Despite these progresses, a general  comprehensive computation of three-point functions involving arbitrary fields is not available yet.\footnote{See ref.\cite{Sotkov:1976xe} for an early attempt.} The knowledge of such correlators is an important ingredient  to extend the recently renewed conformal bootstrap approach \cite{Rattazzi:2008pe} beyond scalar correlators and might as well have applications in the AdS/CFT correspondence and in other contexts.

Aim of this paper is to make a step forward along this direction by computing the most general three-point function in four dimensional (4D) CFTs
between bosonic or fermionic operators in irreducible representations of the Lorentz group. No extra symmetry (like parity) is assumed.
We will achieve this task by extending and generalizing the so called 6D embedding formalism in twistor space as developed by Simmons-Duffin in ref.\cite{SimmonsDuffin:2012uy}.
As in ref.\cite{SimmonsDuffin:2012uy}, we will use an index free notation for the correlators, obtained by saturating indices with auxiliary commuting spinors.
Constraints coming from bosonic or fermionic conserved currents can be simply worked out in this formalism. We will see, generalizing the results found in ref.\cite{Costa:2011mg} for traceless symmetric
operators, that 4D current conservation conditions can be covariantly lifted to 6D only if the conserved operator saturates the unitarity bound.

The extension of our formalism to higher-point functions is in principle straightforward,  but technically complicated.  
Using the OPE, however,  we can at least determine the number of structures appearing in higher-point functions with arbitrary operators by
using our result for three-point functions.  We have in particular  computed in closed form the number of structures appearing in certain  4-point functions with traceless symmetric operators and checked
the consistency of the result using crossing symmetry.

The structure of the paper is as follows. In section 2 we will review the 6D embedding formalism in twistor space in index-free notation and set-up our notation.
We classify and compute all possible three-point functions in section 3. This is the key section of the paper, with eq.(\ref{eq:ff3pf}) being the most important result of this work.  We show in section 4 how the additional constraints imposed by conserved currents are implemented in the 6D twistor space.
The key relations of this section are eqs.(\ref{ConserveddD}) and (\ref{conservationD}).
In order to show the power and simplicity of our formalism, in section 5 we work out explicitly some examples of correlators, with and without conserved operators.
In section 6 we show how our results can be used to compute the number of independent tensor structures of four-point functions and their consistency with crossing symmetry.
We conclude in section 7. Our notation and conventions, as well as useful relations,  are summarized in appendix A, while in appendix B we recall the map between the vector and spinor notation 
for tensor fields.

\section{The 6D Embedding Formalism in Twistor Space in an Index-Free Notation}
\label{sec:6DEmb}

The embedding formalism idea dates back to Dirac \cite{Dirac:1936fq}. It is based on the simple observation that 
the 4D conformal group is isomorphic to $SO(4,2)$, that is  the Lorentz group of a 6D flat space with signature $(--++++)$.
The non-linear action of the conformal group in 4D turns into simple linear Lorentz transformations in 6D.
Hence, by properly extending 4D fields to 6D, one can more easily derive the constraints imposed by the conformal symmetry  on the correlation functions.
The embedding formalism has successfully been used in ordinary space to study correlation functions of traceless symmetric tensors 
\cite{Ferrara:1973yt,Weinberg:2010fx,Costa:2011mg,Costa:2011dw} (see also ref.\cite{Dobrev:1977qv}).  Using the local isomorphism
between $SO(4,2)$ and $SU(2,2)$, the embedding formalism can be reformulated in  twistor space. In this form it has sporadically been used in the literature, mainly in the context of super conformal field theories  (see e.g. refs.\cite{Siegel:1992ic,Siegel:2012di,Goldberger:2011yp,Goldberger:2012xb,Fitzpatrick:2014oza,Khandker:2014mpa}). 
More recently, it has been applied in ref.\cite{SimmonsDuffin:2012uy} to study correlation functions in 4D CFTs.
For completeness,  we briefly review here the embedding formalism in twistor space, essentially following the analysis made in section 5 of ref.\cite{SimmonsDuffin:2012uy}.
We assume that the reader is familiar with basics of CFT.

On $R^{4,2}$, we consider the light-cone defined by  (see appendix \ref{app:Notation} for our notations and conventions)
\be
X^2 = X^M X^N \eta_{MN} = \eta_{\mu \nu} X^\mu X^\nu + X^+ X^- = 0\,.
\label{cone6D}
\ee
We define a projective light-cone by identifying (on the cone) $X^M \cong \lambda X^M$, with $\lambda$ any real non-vanishing constant.
In this way, we have a map between 6D and 4D coordinates.  The standard 4D coordinates  $x^\mu$ should not depend on $\lambda$ and are defined as
\be
x^\mu = \frac{X^\mu}{X^+}\,.
\label{6D4Dcoord}
\ee
It can be shown that conformal transformations acting on $x^\mu$ are mapped to Lorentz transformations acting on the light-cone.

Let us now consider how 4D fields are uplifted to 6D, starting with scalar fields. Let $\phi(x)$ be a 4D primary scalar operator with scaling dimension $\Delta$ and $\Phi(X)$ its corresponding 6D field.
In order to be well defined on the projective cone,  $\Phi(X)$ should be a homogeneous function: $\Phi(\lambda X)= \lambda^{-n} \Phi(X)$, for some $n$.  
A natural identification is
\be
\phi(x) = (X^+)^{n} \Phi(X) \,.
\label{scalar6d}
\ee
It is easy to verify that $n=\Delta$ in eq.(\ref{scalar6d})  to correctly reproduce the conformal transformations of $\phi(x)$.
Let us now consider spin 1/2 primary fermions $\psi_\alpha(x)$ and $\bar\phi^{\dot{\alpha}}(x)$, with scaling dimension $\Delta$.
As shown in ref.\cite{Weinberg:2010fx}, such fields are uplifted to 6D homogeneous  twistors $\Psi_a(X)$ and $\bar\Phi^a(X)$, with degree $n=\Delta-1/2$.
A transversality condition is imposed on the 6D fields, in order to match the number of degrees of freedom:
\be\begin{split}
\overline{\mathbf{X}}^{ab} \Psi_b(X) = \; & 0\,, \\
\bar\Phi^a(X) \mathbf{X}_{ab} = \; & 0 \,,
\end{split}
\label{TransFermions}
\ee
where $\mathbf{X}$ and $\overline{\mathbf{X}}$ are twistor space-time coordinates, defined in eq.(\ref{TwistorCoord}).
By solving eq.(\ref{TransFermions}), we get
\be\begin{split}
 \Psi_a(X) = \;& (X^+)^{-\Delta+1/2}\left(\begin{array}{c}
    \psi_{\alpha}(x)     \\
   -(x_\mu \bar\sigma^\mu)^{\dot\alpha\beta} \psi_{\beta} (x)   
 \end{array}\right)\,, \\
\bar\Phi^a(X)= \;& (X^+)^{-\Delta+1/2}
 \left(\begin{array}{c}
    \bar\phi_{\dot{\beta}}(x)  (x_\mu \bar\sigma^\mu)^{\dot\beta\alpha}      \\
    \bar\phi_{\dot{\alpha}} (x)    
     \end{array}\right)\,.
\end{split}
\label{6DFermionsExp}
\ee
As discussed in ref.\cite{SimmonsDuffin:2012uy}, it is more convenient to embed $\psi_\alpha(x)$ and $\bar\phi^{\dot{\alpha}}(x)$
to twistors $\bar\Psi^a(X)$ and $\Phi_a(X)$, respectively, with degree $n=\Delta+1/2$. In this way, we essentially trade the transversality condition for a gauge redundancy.
A generic solution of eq.(\ref{TransFermions}) is given by $\Psi = \mathbf{X} \bar\Psi$ and $\bar\Phi =  \Phi \overline{\mathbf{X}}$ for some $\bar\Psi$ and $\Phi$, 
since on the cone 
\be
\mathbf{X} \overline{\mathbf{X}}=  \overline{\mathbf{X}}\mathbf{X} =0. 
\label{XXbar}
\ee
We can then equivalently associate $\psi_\alpha(x)$ to a twistor $\bar\Psi^a(X)$, 
and $\bar\phi^{\dot{\alpha}}(x)$ to a twistor $\Phi_a(X)$ as follows:
\be\begin{split}
\psi_\alpha(x) = \; &  (X^+)^{\Delta-1/2} \mathbf{X}_{\alpha a} \bar\Psi^a(X)\,, \\
\bar\phi^{\dot\alpha}(x) = \; & (X^+)^{\Delta-1/2} \overline{\mathbf{X}}^{\dot\alpha a} \Phi_a(X) \,,
\end{split}
\label{6D4DFermionsUp}
\ee
where $\overline{\mathbf{X}}^{\dot\beta b}= \epsilon^{\dot\beta\dot\gamma} \overline{\mathbf{X}}_{\dot\gamma}^{\;\; b}$.
The twistors $\bar\Psi(X)$ and $\Phi(X)$ are subject to an equivalence relation, 
\be\begin{split}
\bar\Psi(X)\sim \; & \bar\Psi(X) +  \overline{\mathbf{X}} V \,, \\
\Phi(X)\sim \; & \Phi(X) + {\mathbf{X}} \overline{W} \,,
\end{split}
\label{Redunda}
\ee
with $V$ and $\overline{W}$ generic twistors. We are now ready to consider a 4D primary spinor-tensor in an arbitrary irreducible representation of the Lorentz group, with scaling dimension $\Delta$:
\be
f_{\alpha_1\ldots \alpha_l}^{\dot{\beta}_1\ldots \dot{\beta}_{\bar l}}(x) \,,
\ee
where dotted and undotted indices are symmetrized. We will denote such a representation as $(l,\bar l)$, namely by the number of undotted and dotted indices that appear.
Hence, a spin 1/2 Weyl fermion will be in the $(1,0)$ or $(0,1)$, a vector in the $(1,1)$, an antisymmetric tensor in the $(2,0)\oplus (0,2)$ and so on.
Generalizing eq.(\ref{6D4DFermionsUp}), we encode $f_{\alpha_1\ldots \alpha_l}^{\dot{\beta}_1\ldots \dot{\beta}_{\bar l}}$ in a 6D multi-twistor field $F^{a_1\ldots a_l}_{b_1\ldots b_{\bar l}}$
of degree $n=\Delta+ (l+\bar l)/2$ as follows:
\be
f_{\alpha_1\ldots \alpha_l}^{\dot{\beta}_1\ldots \dot{\beta}_{\bar l}}(x) =   (X^+)^{\Delta-(l+\bar l)/2}  \mathbf{X}_{\alpha_1 a_1}\ldots  \mathbf{X}_{\alpha_l a_l}  \overline{\mathbf{X}}^{\dot\beta_1 b_1} 
\ldots  \overline{\mathbf{X}}^{\dot\beta_{\bar l} b_{\bar l}} F^{a_1\ldots a_l}_{b_1\ldots b_{\bar l}}(X)\,.
\label{fFrelation}
\ee 
Given the gauge redundancy (\ref{Redunda}) in each index, the 4D field $f$ is uplifted to an equivalence class of 6D fields $F$. Any two fields $F$ and $\hat F= F+ \overline{\mathbf{X}} V$
or $\hat F= F+ {\mathbf{X}} \overline{W}$, for some multi twistors $V$ and $\overline{W}$, are equivalent uplifts of $f$, because of eq.(\ref{XXbar}). There is yet another equivalence class, due again to eq.(\ref{XXbar}).
Twistors of the form $F^{a_1a_2\ldots}_{b_1b_2\ldots} = \delta^{a_1}_{b_1} Z^{a_2\ldots}_{b_2\ldots}$ give a vanishing contribution in eq.(\ref{fFrelation}). Hence, without loss of generality, we can take 
as uplift of $f$ a multi-twistor $F$ with vanishing trace, namely:
\be
\delta^{b_j}_{a_i} F^{a_1\ldots a_l}_{b_1\ldots b_{\bar l}}(X) = 0\,, \ \ \ \forall i=1,\ldots, l, \forall j =1,\ldots,\bar l \,.
\label{notraceF}
\ee
It is very useful to use an index-free notation by defining
\be
f(x,s,\bar s) \equiv f_{\alpha_1\ldots \alpha_l}^{\dot{\beta}_1\ldots \dot{\beta}_{\bar l}}(x) s^{\alpha_1} \ldots s^{\alpha_l} \bar s_{\dot{\beta}_1}\ldots \bar s_{\dot{\beta}_{\bar l}}\,,
\label{findexfree}
\ee
where $s^\alpha$ and $\bar s_{\dot\beta}$ are auxiliary (commuting and independent) spinors. Similarly, we define
\be
F(X,S,\bar S) \equiv  F^{a_1\ldots a_l}_{b_1\ldots b_{\bar l}}(X)\ S_{a_1} \ldots S_{a_l} \bar S^{b_1} \ldots \bar S^{b_{\bar l}}
\label{Findexfree}
\ee
in terms of auxiliary (again commuting and independent) twistors $S_a$ and $\bar S^a$. 
We demand that, upon projection down to 4D,
\be
 (X^+)^{\Delta+(l+\bar l)/2} F(X,S,\bar S)   \stackrel{4D}{\longrightarrow}  f(x,s,\bar s) \,.
\label{Proje46D}
\ee
Consistency of eqs.(\ref{fFrelation}), (\ref{findexfree})-(\ref{Proje46D}) implies that
\be
S_a \stackrel{4D}{\longrightarrow }  s^\alpha \frac{\mathbf{X}_{\alpha a}}{X^+}\,, \ \ \ \ \bar S^a  \stackrel{4D}{\longrightarrow}   \bar s_{\dot \beta}\frac{ \overline{\mathbf{X}}^{\dot\beta a}}{X^+} \,.
\label{Ssaux}
\ee
From eq.(\ref{Ssaux}) we also deduce
\be\label{eq:gauge2}
\overline{\mathbf{X}}^{ab} S_b \stackrel{4D}{\longrightarrow } 0\,, \ \ \ \bar S^b  \mathbf{X}_{b a}  \stackrel{4D}{\longrightarrow }  0\,, \ \ \ \bar S^a S_a  \stackrel{4D}{\longrightarrow }  0\,,
\ee
consistently with the gauge redundancies we have in choosing $F$. Given a 6D multi-twistor field $F$, the corresponding 4D field $f$ is explicitly given by
\be
 f_{\alpha_1\ldots \alpha_l}^{\dot{\beta}_1\ldots \dot{\beta}_{\bar l}}(x) = \frac{ (X^+)^{\Delta-\frac{l+\bar l}{2}} }{l!\bar l!} \Big( {\mathbf{X}}\frac{\partial}{\partial S}\Big)_{\alpha_1} \ldots  \Big( {\mathbf{X}}\frac{\partial}{\partial S}\Big)_{\alpha_l} \Big( {\overline{\mathbf{X}}}\frac{\partial}{\partial \bar S}\Big)^{\dot\beta_1}\ldots \Big( {\overline{\mathbf{X}}}\frac{\partial}{\partial \bar S}\Big)^{\dot\beta_{\bar l}}
F\Big(X,S,\bar S\Big) \,.
  \label{f4dExp}
\ee
It is useful to compare the index-free notation introduced here with the one introduced in ref.\cite{Costa:2011mg} for symmetric traceless tensors in terms of polynomials in auxiliary variables $z^\mu$ and $Z^M$.
Recall that in vector notation, a 4D symmetric traceless tensor $t_{\mu_1\ldots \mu_l}$ can be embedded in a 6D tensor $T_{M_1\ldots M_l}$ by means of the relation
\be
t_{\mu_1\ldots \mu_l}  = (X^+)^{\Delta-l} \frac{\partial X^{M_1}}{\partial x^{\mu_1}}\ldots   \frac{\partial X^{M_l}}{\partial x^{\mu_l}}T_{M_1\ldots M_l} \,,
\ee
where $T_{M_1\ldots M_l}$ is symmetric traceless in 6D, homogeneous of degree $\Delta$, as well as transverse: $X^{M_1} T_{M_1M_2\ldots M_l}=0$.
In ref.\cite{Costa:2011mg}, 4D and 6D fields are encoded in the polynomials
\be\begin{split}
t(x,z) \; = \; & t_{\mu_1\ldots \mu_n} z^{\mu_1}\ldots z^{\mu_n}\,, \\
T(X,Z) \; = \; & T_{M_1\ldots M_n} Z^{M_1}\ldots Z^{M_n}  \,,
\end{split}
\ee
where in Minkowski space $z_\mu$ is a light-cone vector, $z_\mu z^\mu=0$.
A null vector can always be written as a product of two spinors:
\be
z^\mu = \sigma^\mu_{\alpha\dot\beta} s^\alpha\bar s^{\dot\beta} \,.
\label{zmuss}
\ee
Given the relation  (\ref{ChangebasisTrace}) between symmetric traceless tensors written in vector and spinor notation, the spinors $s^\alpha$ and $\bar s^{\dot\alpha}$ appearing in eq.(\ref{zmuss}) are 
exactly the ones defined in eq.(\ref{findexfree}). On the contrary, there is not a simple relation between the 6D coordinates $Z^A$ and the 6D twistors $S_a$ and $\bar S^a$.

\section{Three-Point Functions}
\label{sec:3pointFun}

The goal of this section is to classify and compute the most general three-point function in a 4D CFT  using the 6D embedding formalism reviewed in section \ref{sec:6DEmb}, essentially completing
the program that was outlined in ref.\cite{SimmonsDuffin:2012uy}, where this formalism was first proposed and used. Although some of the results of this section were already obtained in ref.\cite{SimmonsDuffin:2012uy}, for 
the clarity of the presentation and for completeness, they will be reported here in a more systematic framework.

Three-point functions in a CFT are completely fixed by the conformal symmetry,  up to a set of constants. 
Let us denote by $F_i=F_i(X_i,S_i,\bar{S}_i)$ the  index-free 6D multi tensor field corresponding to some $(l_i,\bar l_i)$ 4D tensor field $f_i$. 
An arbitrary three-point function can schematically be written as 
\be\label{eq:3cf}
\langle F_1 F_2 F_3 \rangle =\mathcal{K}\;\sum_{s=1}^{N_3}\lambda_s \mathcal{T}_s,
\ee
where $\mathcal{K}$ is a kinematic factor which  depend on the scaling dimension and spin of the external fields, $\mathcal{T}_s$ are dimensionless (i.e. homogeneous with degree zero) 
$SU(2,2)$ invariant tensor structures which encode the Lorentz structure of the fields and $\lambda_s$ are constants. The index $s$ runs over all the possible different independent tensor structures compatible with conformal invariance.

Let us start with the kinematic factor
\be\label{eq:kinematicfactor1}
\mathcal{K}=\frac{1}{X_{12}^{a_{12}} X_{13}^{a_{13}} X_{23}^{a_{23}}},
\ee
where we use a 6D short-hand notation
\be 
X_{ij}\equiv X_i\cdot X_j\,.
\ee
The coefficients $a_{ij}$ are determined by matching the scaling dimension of both sides of eq.(\ref{eq:3cf}):
\be
a_{ij} =\frac{1}{2}\Bigg(\Delta_{ijk}+\frac{(l_i+\bar{l}_i)+(l_j+\bar{l}_j)-(l_k+\bar{l}_k)}{2}\Bigg)  \,,  \ \ \ i\neq j\neq k \,,
\ee
where we have defined
\be
\Delta_{ijk} \equiv \Delta_i+\Delta_j-\Delta_k = \Delta_{jik} \,.
\ee
Finding the tensor structures $\mathcal{T}_s$ is a much less trivial problem. Any $\mathcal{T}_s$ will be a product of some fundamental $SU(2,2)$ invariant building blocks that are to be determined.

\subsection{Invariant Building Blocks}
\label{subsec:BB}

The fundamental group-theoretical objects carrying $SU(2,2)$ indices, which should  eventually be combined with the auxiliary twistors $S_a$ and $\bar S^b$ to form $SU(2,2)$ invariants, are obtained as products of 
\be\label{eq:basictensstruct}
\delta^a_b,\;\varepsilon_{abcd},\;\varepsilon^{abcd},\;\mathbf{X}_{ab},\;\overline{\mathbf{X}}^{ab}.
\ee
Let us first focus on the $\varepsilon$ tensors. Their contraction with any other object in eq.(\ref{eq:basictensstruct}) does not give any new structures, because they reduce to a sum of already existing elements in eq.(\ref{eq:basictensstruct}), for example:
\begin{align}
\varepsilon^{abcd}\varepsilon_{aefg} &= \delta^b_e \delta^c_f \delta^d_g-\delta^b_e \delta^c_g \delta^d_f
-\delta^b_f \delta^c_e \delta^d_g+\delta^b_f \delta^c_g \delta^d_e
+\delta^b_g \delta^c_e \delta^d_f-\delta^b_g \delta^c_f \delta^d_e,\\
\varepsilon^{abcd}\mathbf{X}_{ae}    &= -\delta^b_e\overline{\mathbf{X}}^{cd}+\delta^c_e\overline{\mathbf{X}}^{bd}
-\delta^d_e\overline{\mathbf{X}}^{bc}.
\end{align} 
Actually, for three-point functions the $\varepsilon$-symbols drop from the discussion completely. It can be seen using the index-free formalism where $\varepsilon$ is encoded into $\varepsilon_{abcd}\bar{S}_i^a\bar{S}_j^b\bar{S}_k^c\bar{S}_l^d$, which vanishes unless $i\neq j\neq k\neq l$. Thus, the tensors $\varepsilon$ become relevant starting from the four-point functions.
The fundamental group-theoretical objects can be grouped into three sets
\be \Big\lbrace \delta_a^b,\; [\mathbf{X}_i\overline{\mathbf{{X}}}_j]_a^{\;\,b},\; [\mathbf{X}_i\overline{\mathbf{X}}_j \mathbf{X}_k\overline{\mathbf{X}}_l]_a^{\;\,b},
\ldots\Big\rbrace , \Big\lbrace [\overline{\mathbf{X}}_i]^{ab},[\overline{\mathbf{X}}_i\mathbf{X}_j\overline{\mathbf{X}}_k]^{ab},
\ldots\Big\rbrace ,
\Big\lbrace [\mathbf{X}_i]_{ab},[\mathbf{X}_i\overline{\mathbf{X}}_j\mathbf{X}_k]_{ab},
\ldots\Big \rbrace.
\label{eq:set}
\ee

Multiplying these objects by auxiliary twistors $S$ and $\bar{S}$ will give us the SU(2,2)  invariant building blocks needed to characterize the three-point (or any other $n$-point) function. 
They are not all independent, given the relations (\ref{XXbar}),  (\ref{eq:gauge2}) and (\ref{eq:Xperm}). 

Let us first determine the general form of two-point functions $\langle F_1 F_2\rangle$.
It is clear in this case that the only non-vanishing independent SU(2,2) invariant is obtained by contracting one twistor $\bar S_{1}$ with $S_{2}$ or viceversa.
The form of the two-point function is uniquely determined:
\be
\langle F_1(X_1,S_1,\bar S_1) F_2(X_2,S_2,\bar S_2) \rangle = c X_{12}^{-\big(\Delta_1+\frac{l_1+\bar l_1}2\big)} I_{21}^{l_1} I_{12}^{\bar l_1} \delta_{l_1,\bar l_2} \delta_{l_2,\bar l_1} \delta_{\Delta_1, \Delta_2} \,,
\ee
where $c$ is a normalization factor and we have defined the SU(2,2) invariant
\be
I_{ij}   \equiv \bar S_i S_j  \,.
\label{eq:invar1}
\ee
For three-point functions three more invariants arise:
\begin{align}
\label{eq:invar2}
K_{i,jk}            &\equiv N_{i,jk} S_j \overline{\mathbf{X}}_i S_k \,, \\
\label{eq:invar3}
\overline{K}_{i,jk} &\equiv N_{i,jk} \bar S_j \mathbf{X}_i \bar S_k \,,\\
\label{eq:invar4}
J_{i,jk}            &\equiv  N_{jk} \bar S_i  \mathbf{X}_j \overline{\mathbf{X}}_k S_i \,.
\end{align}
The normalization factors
\be
N_{jk}   \equiv \frac{1}{X_{jk}}\,, \ \ \ 
N_{i,jk}  \equiv \sqrt{\frac{X_{jk}}{X_{ij}X_{ik}}}\,,
\label{eq:Ninva}
\ee
are introduced to make the $SU(2,2)$ invariants in eqs.(\ref{eq:invar1})-(\ref{eq:invar4}) dimensionless and well-defined on the 6D light-cone.\footnote{Notice the different normalization and slight different index notation in the definition of the invariants $I$, $K$, $\overline{K}$ and $J$ with respect to the ones defined in ref.\cite{SimmonsDuffin:2012uy}.} 
Notice that in eqs.(\ref{eq:invar2})-(\ref{eq:invar4}) $i\neq j\neq k$ and indices are not summed.
The invariants (\ref{eq:invar2})-(\ref{eq:invar4}) are all anti-symmetric in the two indices after the comma:
\begin{equation}\label{eq:invarsymm}
K_{i,jk}=-K_{i,kj},\;\;\overline K_{i,jk}=-\overline K_{i,kj},\;\;J_{i,jk}=-J_{i,kj},
\end{equation}
due to the anti-symmetry of $\mathbf{X}$, $\overline{\mathbf{X}}$ and the relations (\ref{eq:gauge2}), (\ref{eq:Xperm}).

Every other $SU(2,2)$ invariant object obtained from eq.(\ref{eq:set}) can be written in terms of different combinations of $I_{ij},K_{i,jk},\overline{K}_{i,jk}$ and $J_{i,jk}$. 
Using eqs.(\ref{eq:invar1})-(\ref{eq:invar4}), the most general tensor structure can be written as follows:
\be \label{eq:generaltensstruct}
\mathcal{T}_s=\lambda_s\, I_{12}^{m_{12}} I_{21}^{m_{21}} I_{13}^{m_{13}} I_{31}^{m_{31}} I_{23}^{m_{23}} I_{32}^{m_{32}} K_{1,23}^{k_1} K_{2,13}^{k_2} K_{3,12}^{k_3} \overline{K}_{1,23}^{\bar k_1} \overline{K}_{2,13}^{\bar k_2} \overline{K}_{3,12}^{\bar k_3}  J_{1,23}^{j_1} J_{2,13}^{j_2} J_{3,12}^{j_3},
\ee
where $m_{ij}$, $k_i$, $\bar k_i$ and $j_i$ are a set of non-negative integers.  Matching the powers of $S_i$ and $\bar{S}_j$ in both sides of eq.(\ref{eq:3cf}) gives us six constraints: 
\be \label{eq:thesystem}
\begin{cases}
l_1       &= m_{21} + m_{31} +  0  + k_2 + k_3 + j_1  \\
l_2       &= m_{12} + m_{32} + k_1 +  0  + k_3 + j_2 \\
l_3       &= m_{13} + m_{23} + k_1 + k_2 +  0  + j_3 \\
\bar{l}_1 &= m_{12} + m_{13} +  0  + \bar k_2 + \bar k_3 + j_1 \\
\bar{l}_2 &= m_{21} + m_{23} + \bar k_1 +  0  + \bar k_3 + j_2 \\
\bar{l}_3 &= m_{31} + m_{32} + \bar k_1 + \bar k_2 +  0  + j_3\,. 
\end{cases}
\ee
This would have completed the classification of the three-point functions if the  tensor structures $\mathcal{T}_s$ were all linearly independent, but they are not, and hence a more refined analysis is necessary.

\subsection{Relations between Invariants}

The dependence of the structures (\ref{eq:generaltensstruct}) has its roots in a set of identities among the twistors $S_i$ and the coordinates $\mathbf{X}_j$, when $i=j$.
Recall that on the 6D light-cone $\mathbf{X}$ can be written in terms of auxiliary twistors $V$ and $W$:
\be\label{eq:usefulform1}
\mathbf{X}_{ab}=V_aW_b-V_bW_a.
\ee
The twistor $S$ can also be rewritten in an analogous manner. We solve eq.(\ref{eq:gauge2}) by $S_a=\mathbf{X}_{ab}\bar{T}^b$ for some $\bar{T}^b$ and then 
we use eq.(\ref{eq:usefulform1}) to get
\be\label{eq:usefulform2}
S_{a}=\alpha V_a+\beta W_a\,,
\ee
with $\alpha = \bar T W$, $\beta=-\bar T V$. Using eqs.(\ref{eq:usefulform1}) and (\ref{eq:usefulform2}) it is immediate to verify the identities
\begin{align}
S_a\mathbf{X}_{bc} + S_b\mathbf{X}_{ca}+S_c\mathbf{X}_{ab} = &  \; 0\, ,
\label{Jset1} \\
\mathbf{X}_{ab}\mathbf{X}_{cd} + \mathbf{X}_{ca}\mathbf{X}_{bd}+\mathbf{X}_{bc}\mathbf{X}_{ad}= &\;  0\,.
\label{Jset2}
\end{align}
Analogous relations apply for the dual twistors $\bar S$ and $\overline{\mathbf{X}}$. We have not found identities involving more $S$'s or $\mathbf{X}$'s that do not boil down 
to eqs.(\ref{Jset1}) and (\ref{Jset2}).  Applying eqs.(\ref{Jset1}) and (\ref{Jset2}) (actually it is enough to use only eq.(\ref{Jset1})) to bi-products of invariants we get the following relations
(no sum over indices):
\begin{align}
          K_{{j},ik}\overline{K}_{{i},jk} & =2I_{ki}I_{jk}-I_{ji}J_{k,ij}\,,
        \label{eq:setrel1}  \\
      J_{{i},jk}J_{{j},ik} & =2\big(2I_{ij}I_{ji}+K_{{k},ij}\overline{K}_{{k},ij}\big)\,,
      \label{eq:setrel2} \\
  J_{{j},ik}K_{{j},ik}       & =2\big(-I_{ji}K_{i,jk}+I_{jk}K_{k,ij}\big)\,,
     \label{eq:setrel3}   \\
  J_{{j},ik}\overline{K}_{{j},ik} & =2\big(-I_{ij}\overline{K}_{i,kj}-I_{kj}\overline{K}_{k,ij}\big) \,.
      \label{eq:setrel4}
\end{align}
We have verified that higher order relations involving more than 2 invariants always arise as the composition of the
relations (\ref{eq:setrel1})-(\ref{eq:setrel4}). This is expected, since the fundamental identities (\ref{Jset1}) and (\ref{Jset2}) involve only two tensors.
A particularly useful third-order relation is 
\be\label{eq:setrel5}
J_{1,23}J_{2,13}J_{3,12}= 8\,\big(I_{21}I_{13}I_{32}-I_{12}I_{31}I_{23}\big) + 4\,\big(  I_{23}I_{32}J_{1,23} - I_{13}I_{31}J_{2,13} + I_{12}I_{21}J_{3,12}  \big),
\ee
which is obtained by applying, in order, eqs.(\ref{eq:setrel2}), (\ref{eq:setrel4}) and (\ref{eq:setrel1}).
The relations (\ref{eq:setrel1})-(\ref{eq:setrel5}) have been originally obtained in ref.\cite{SimmonsDuffin:2012uy}, though it was not clear there whether additional relations were possible. 

Combining eqs.(\ref{eq:setrel1}) and (\ref{eq:setrel2}),  we see that a product of any $K$ and $\overline{K}$ can be reduced to a combination of $I$'s and $J$'s. 
Thus, we obtain the first constraint on the integers $k_i$ and $\bar k_i$ appearing in eq.(\ref{eq:generaltensstruct}):
\be\label{eq:constrpowers1}
k_1=k_2=k_3=0\;\;\;\mathbf{or}\;\;\; \bar k_1=  \bar k_2=  \bar k_3= 0 \,.
\ee
In other words, we can always choose a basis of invariants $ \mathcal{T}_s$ where $K$'s and $\overline{K}$'s never appear together.
Next we can apply eq.(\ref{eq:setrel5}) successively. At each step the tensor structure splits into five ones, each time with a reduced number of $J$'s. We keep applying eq.(\ref{eq:setrel5})
until the initial tensor structure is written as a sum of tensor structures where all have at least one value of $j_1$, $j_2$, or $j_3$ equal to zero.
Thus we get the second constraint in eq.(\ref{eq:generaltensstruct}):
\be\label{eq:constrpowers2} 
j_1=0\;\;\;\mathbf{or}\;\;\; j_2=0\;\;\;\mathbf{or}\;\;\; j_3=0.
\ee
The last step is to apply eq.(\ref{eq:setrel4})  (for $k_{1,2,3}=0$) or eq.(\ref{eq:setrel3}) (for $\overline k_{1,2,3}=0$), so that  products of the form $K_{i,..}J_{i,..}$ or $\overline{K}_{i,..}J_{i,..}$can be rewritten using only $K$'s or $\overline{K}$'s of a different type. It is not difficult to convince oneself that this boils down to the following further constraints on eq.(\ref{eq:generaltensstruct}): 
\be 
\left\{\begin{array}{c}
\label{eq:constrpowers3} 
k_1=0\;\;\;\mathbf{or}\;\;\; j_1=0     \\
k_2=0\;\;\;\mathbf{or}\;\;\; j_2=0     \\
k_3=0\;\;\;\mathbf{or}\;\;\;j_3=0    
\end{array} \right. \ \ \ \ 
\left\{\begin{array}{c}
\bar k_1=0\;\;\;\mathbf{or}\;\;\; j_1=0     \\
\bar k_2=0\;\;\;\mathbf{or}\;\;\; j_2=0     \\
\bar k_3=0\;\;\;\mathbf{or}\;\;\; j_3=0     \,.
\end{array} \right.
\ee

\subsection{Final Classification of Tensor Structures and Further Considerations}

There are no further relations to be imposed so we can finally state the main result of this paper. 

{\it The most general three-point function $\langle F_1 F_2 F_3 \rangle$ can be written as
 \be\label{eq:ff3pf}
\langle F_1 F_2 F_3 \rangle= \mathcal{K}\;\sum_{s=1}^{N_3} \lambda_s  \Big(\prod_{i\neq j=1}^3 I_{ij}^{m_{ij}} \Big) K_{1,23}^{k_1} K_{2,13}^{k_2} K_{3,12}^{k_3} \overline{K}_{1,23}^{\bar k_1} \overline{K}_{2,13}^{\bar k_2} \overline{K}_{3,12}^{\bar k_3}  J_{1,23}^{j_1} J_{2,13}^{j_2} J_{3,12}^{j_3},
\ee
where $\mathcal{K}$ is given by eq.(\ref{eq:kinematicfactor1}) and $s$ runs over all the independent tensor structures. These are given by the set of non-negative exponents $m_{ij}$, $k_i$, $\bar k_i$ and $j_i$ solution of eq.(\ref{eq:thesystem}) and subjected to the constraints (\ref{eq:constrpowers1}), (\ref{eq:constrpowers2}) and (\ref{eq:constrpowers3}).}
The latter require that, modulo the $I_{ij}$ invariants, at most three more invariants can be present in each tensor structure. We can have i) 2 $J$'s, ii) 3 $K$'s, iii) 2 $K$'s and 1 $J$,  iv) 1 $K$ and 2$J$'s,
v) 3 $\overline{K}$'s, vi) 2 $\overline{K}$'s and 1 $J$, vii) 1 $\overline{K}$ and 2 $J$'s.  

Let us discuss some implications of eq.(\ref{eq:ff3pf}). It is useful to define
\be\label{eq:dl}
\Delta l\equiv l_1+l_2+l_3-(\bar l_1+\bar l_2+\bar l_3)\,.
\ee 
Using the system (\ref{eq:thesystem}), we immediately get 
\bea
\Delta l & = &  2(k_1+k_2+k_3-\bar k_1-\bar k_2-\bar k_3)\,, 
\label{eq:criterion1} \\
k_1+k_2+k_3 &&\!\!\!\! \leq  min(l_1+l_2,l_1+ l_3 ,l_2+l_3 )\,,  \ \ \ \bar k_1+\bar k_2+\bar k_3  \leq min(\bar l_1+\bar l_2,\bar l_1+\bar l_3 ,\bar l_2+\bar l_3 ) \,, \nn
\eea
and hence
\be
-2\min  (\bar l_1+\bar l_2,\,\bar l_1+\bar l_3,\,\bar l_2+\bar l_3) \leq  \Delta l \leq
2\min (l_1+l_2,\,l_1+l_3,\,l_2+l_3)\,.
\label{eq:criterion2}
\ee
These are the conditions for the 4D three-point function  $\langle f_1 f_2 f_3 \rangle$ to be non-vanishing. They exactly match the findings of ref.\cite{Mack:1976pa}.
Indeed, in that paper it was found that the 3-point function $\langle f_1 f_2 f_3 \rangle$, with  $f_i$ primary fields in the $(l_i,\bar{l}_i)$ representations of $SL(2,C)$, is non-vanishing
if  the decomposition of the tensor product $({l_1},\bar{l}_1) \otimes ({l_2},\bar{l}_2) \otimes ({l_3},\bar{l}_3)$
contains a traceless-symmetric representation $(l,l)$. Then we have
\be\label{eq:decompos}
({l_1},\bar{l}_1) \otimes ({l_2},\bar{l}_2) \otimes ({l_3},\bar{l}_3) =\sum_{m=0}^{min( l_1, l_2)}\sum_{\bar m=0}^{min(\bar l_1,\bar l_2)}\ \sum_{p=0}^{p_m}\sum_{\bar p=0}^{\bar p_m}(l_1+l_2+l_3-2m-2p,\bar l_1+\bar l_2+\bar l_3-2\bar m -2\bar p ),
\ee
where $p_m =min(l_1+l_2-2m,l_3) $, $\bar p_m = min(\bar l_1+\bar l_2-2\bar m,\bar l_3)$, and the indices of summation are subjected to the following constraints
\be\label{eq:lconstr}
m+p \leq min(l_1+l_2,l_1+ l_3 ,l_2+l_3 ),\;\;\bar m +\bar p \leq min(\bar l_1+\bar l_2,\bar l_1+\bar l_3 ,\bar l_2+\bar l_3 ).
\ee
Demanding that a term of the form $(l,l)$ appears in the r.h.s. of eq.(\ref{eq:decompos}) implies 
\be\label{eq:justl}
\Delta l=2(m+p-\bar m-\bar p),
\ee
where $\Delta l$ is defined in eq.(\ref{eq:dl}). We then see that eqs.(\ref{eq:lconstr}) and (\ref{eq:justl}) exactly correspond to eqs.(\ref{eq:criterion1}), with the identification 
$m+p \rightarrow  k_1+k_2+k_3$, $\bar m+\bar p \rightarrow  \bar k_1+\bar k_2+\bar k_3$.

The master formula (\ref{eq:ff3pf}) computes the most general three-point function compatible with conformal symmetry. Invariance under parity transformations, in particular, is {\it not} assumed.
It should be obvious that additional symmetries (like exchange symmetries with identical operators or conserved operators) put further constrains on the form of the 3-point function. 
We will consider in more detail parity in subsection \ref{subsec:parity} and conserved operators in section \ref{sec:CurrentCons}.

For any given correlator, the explicit form and the number $N_3$ of independent tensor structures is easily determined with a numerical algorithm.
In general, $N_3=N_3(l_1,\bar l_1,l_2,\bar l_2,l_3,\bar l_3)$ and it is a laborius task (which we have not tried to do) to find its analytic expression for any correlator.
However, we have been able to get a simple formula for the correlations involving two traceless-symmetric tensors ($l_1,l_1)$, $(l_2,l_2)$ and an arbitrary ($l_3,\bar l_3$) field. 
The number of independent structures is found to be
\be\begin{split}
N_3(l_1,l_1,l_2,l_2,l_3,\bar l_3) & =  \; 1+l_2\,(l_1+1)^2-\frac{1}{3}\,l_1\,(l_1^2-4)+\frac{1}{24}\,|\Delta l|\,(\Delta l^2-4)-\frac{1}{4}\,\Delta l^2\, (l_1+1)\\
&+\frac{1}{6}\,q\,(q^2-1) -\frac{1}{3}\,m_1\,(m_1-1)\,(m_1-2)-\frac{1}{3}\,m_2\,(m_2-1)\,(m_2-2),
\end{split}
\label{eq:counting}\ee
where $\Delta l = l_3-\bar{l}_3$ and
\be 
q=\max(0,\frac{1}{2}\,|\Delta l|+l_1-l_2),\;\;m_1=\max(0,\,\frac{1}{2}|\Delta l|-l_1),\;\;m_2=\max(0,\,\frac{1}{2}|\Delta l|-l_2).
\ee
The domain of validity of this formula is $l_1\leq l_2$, $l_1+l_2-\frac{1}{2}|\Delta l| \leq \min(l_3,\bar{l}_3)$ and $|\Delta l|\leq 2(l_1+l_2)$. When $\bar l_3=l_3$ eq.(\ref{eq:counting}) agrees  with the analytic counting of independent tensor structures performed in ref.\cite{Costa:2011mg}.\footnote{When matching our result with ref.\cite{Costa:2011mg} one should not forget that eq.(\ref{eq:counting}) counts both parity-even and parity-odd structures.}

In principle the classification performed here for three-point functions can be extended to four (or higher) point functions, although its complexity  rapidly grows.
There are 64 SU(2,2) invariant building blocks  (compared to 15 for three-point functions) 
and many  relations among bi-products of invariants for four-point functions. For this reason we have not attempted to make a general classification 
of correlators with more than three fields.

\subsection{6D to 4D Dictionary}
\label{app:useful}

The projection from the 6D to the 4D index-free forms is extremely easy.  Given a 6D three-point function, we just need to project the invariants $I,K,\bar{K},J$  using eq.(\ref{Proje46D}).
We have
\bea\label{IijExp}
I_{ij}&  \stackrel{4D}{\longrightarrow } & \frac{1}{X_i^+X_j^+}\bar{s}_{i\dot\alpha} (\overline{\mathbf{X}}_i  \mathbf{X}_j)^{\dot\alpha}_\alpha s_j^{\alpha} = \bar{s}_{i\dot\alpha}(x_{ij}\cdot \sigma\epsilon)_{\alpha}^{\dot\alpha} \,s_j^{\alpha}, \\
K_{k,ij}&   \stackrel{4D}{\longrightarrow }  & \frac{ N_{k,ij}}{X_i^+ X_j^+}
s_i^\alpha s_j^\beta(\mathbf{X}_i\overline{\mathbf{X}}_k \mathbf{X}_j)_{\alpha\beta}  \nn \\
&  = & \frac{i}{\sqrt{2}}\,\frac{|x_{ij}|}{|x_{ik}||x_{jk}|}s_i^\alpha s_j^\beta 
\bigg(
(x_{ik}^2+x_{jk}^2-x_{ij}^2)\epsilon_{\alpha\beta}+4x_{ik}^\mu x_{kj}^\nu(\sigma_{\mu\nu}\epsilon)_{\alpha\beta} 
\bigg) \,, \\
\overline K_{k,ij}&   \stackrel{4D}{\longrightarrow }  & \frac{ N_{k,ij}}{X_i^+ X_j^+}\bar{s}_{i\dot\alpha}\bar{s}_{j\dot\beta}(\overline{\mathbf{X}}_i  \mathbf{X}_k \overline{\mathbf{X}}_j )^{\dot\alpha\dot\beta} \nn  \\
& = & \frac{i}{\sqrt{2}}\,\frac{|x_{ij}|}{|x_{ik}||x_{jk}|} \bar{s}_{i\dot\alpha}\bar{s}_{j\dot\beta}
\bigg(
(x_{ik}^2+x_{jk}^2-x_{ij}^2)\epsilon^{\dot\alpha\dot\beta}+4x_{ik}^\mu x_{kj}^\nu
(\bar\sigma_{\mu\nu}\epsilon)^{\dot\alpha\dot\beta}
\bigg) \,, \\
J_{k,ij}&  \stackrel{4D}{\longrightarrow } & \frac{N_{ij}}{(X_k^+)^2} \bar{s}_{k\dot\alpha} (\overline{\mathbf{X}}_k  \mathbf{X}_i\overline{\mathbf{X}}_j \mathbf{X}_k)^{\dot\alpha}_{\alpha} s_k^{\alpha} = 2\frac{x_{ik}^2x_{kj}^2}{x_{ij}^2} s_k^{\alpha} (Z_{k,ij}\cdot\sigma\epsilon)_{\alpha}^{\dot\alpha}\bar{s}_{k\dot\alpha} \,,
\label{JijExp}
\eea
where
\begin{equation}
Z_{k,\,ij}^\mu\equiv \frac{x_{ki}^\mu}{x_{ki}^2}-\frac{x_{kj}^\mu}{x_{kj}^2},\;\;Z_{k,\,ij}^\mu=-Z_{k,\,ji}^\mu\,.
\label{Zkmu}
\end{equation}  
Explicit 4D correlation functions with indices are obtained by removing the auxiliary spinors $s_i$ and $\bar s_i$ through derivatives, as described in eq.(\ref{f4dExp}).

\subsection{Transformations under 4D Parity}
\label{subsec:parity}

Under the 4D parity transformation ${\cal P}:(x^0,\vec{x})\rightarrow (x^0,-\vec{x})$, a 4D field in the $(l,\bar l)$ representation of the Lorentz group is mapped to a field in the complex conjugate representation $(\bar l,l)$.
We parametrize the transformation as follows:
\be
 f_{\alpha_1\ldots \alpha_l}^{\dot{\beta}_1\ldots \dot{\beta}_{\bar l}}(x) \stackrel{{\cal P}}{\longrightarrow} \eta (-)^{\frac{l+\bar l}{2}}   f^{\dot\alpha_1\ldots \dot\alpha_l}_{\beta_1\ldots \beta_{\bar l}}(\tilde x) \,,
\label{parity}
\ee
where $\eta$ is the intrinsic parity of the field and $\tilde x$ is the parity transformed coordinate. Applying parity twice gives
\be
\eta \, \eta_c (-)^{l+\bar l} = 1\,,
\ee
where $\eta_c$ is the intrinsic parity of the conjugate field. We then see that $\eta \eta_c = +1$ for bosonic operators and $\eta \eta_c = -1$ for fermionic ones. 
Under parity, in particular,  we have
\be
(x\cdot \sigma \epsilon)_\alpha^{\dot\beta}\stackrel{{\cal P}}{\longleftrightarrow} - (x\cdot\bar \sigma \epsilon)_\beta^{\dot\alpha}\,, \ \ \ \ 
\epsilon_{\alpha\beta} \stackrel{{\cal P}}{\longleftrightarrow} - \epsilon^{\dot\alpha\dot\beta}\,, \ \ \ 
x^\mu y^\nu (\sigma_{\mu\nu}\epsilon)_{\alpha\beta} \stackrel{{\cal P}}{\longleftrightarrow} -x^\mu y^\nu (\bar\sigma_{\mu\nu}\epsilon)^{\dot\alpha\dot\beta}  \,.
\label{paritytrans}
\ee
We can see how parity acts on the 6D invariants (\ref{eq:invar1})-(\ref{eq:invar4}) by using their 4D expressions (\ref{IijExp})-(\ref{JijExp}) on the null cone and eqs.(\ref{paritytrans}).
We get
\be\begin{split}
I_{\ij} \stackrel{{\cal P}}{\longrightarrow} \;\; & - I_{ji} \,,\\
K_{i,jk}\stackrel{{\cal P}}{\longrightarrow} \;\;  & + \overline{K}_{i,jk} \,, \\ 
\overline{K}_{i,jk}\stackrel{{\cal P}}{\longrightarrow} \;\;  & + K_{i,jk} \,, \\
 J_{i,jk}\stackrel{{\cal P}}{\longrightarrow} \;\;  & + J_{i,jk} \,.
 \label{6Dparity}
\end{split}
\ee
In general, parity maps  correlators of fields into correlators of their complex conjugate fields. Imposing parity in a CFT implies that for each primary field $(l,\bar l)$ there must exist
its conjugate one $(\bar l,l)$, and the constants entering in their correlators are related. Of course, we can also have correlators that are mapped to themselves under parity.
Since $\Delta l\rightarrow -\Delta l$ under parity, where $\Delta l$ is defined in eq.(\ref{eq:dl}), 
such correlators should have $\Delta l=0$. Due to eqs.(\ref{eq:criterion2}) and (\ref{eq:constrpowers1}),  the structures $K$ and $\overline{K}$ cannot enter in correlators with $\Delta l=0$, which 
depend only on the invariants $I_{ij}$ and $J_{i,jk}$.
For correlators that are mapped to themselves under parity, one has to take linear combinations of the tensor structures appearing 
in eq.(\ref{eq:ff3pf}) that are even or odd under parity, according to the transformation rules for $I$'s and $J$'s in eq.(\ref{6Dparity}). 
Depending on the intrinsic parity of the product of the fields entering the correlator, the coefficients multiplying the parity even or parity  odd structures should then be set to zero if parity is conserved.

A particular relevant class of correlators that are mapped to themselves under parity are those involving symmetric traceless tensors only. In this case we have verified that eqs.(\ref{6Dparity}) lead to the correct number of parity even and parity odd structures as separately computed in ref.\cite{Costa:2011mg}.

\section{Conserved Operators}
\label{sec:CurrentCons}

Primary tensor fields whose scaling dimension $\Delta$  saturates the unitarity bound  \cite{Mack:1975je} (see also ref.\cite{Minwalla:1997ka} for a generalization
to $D\neq 4$ space-time dimensions)
\be
\Delta \geq \frac{l+\bar l}2 +2 \,, \ \ l\neq 0 \;\; \mathbf{and} \ \ \bar l\neq 0\,,
\label{UnitarityBound}
\ee
are conserved. Three-point functions with conserved operators are subject to further constraints which will be analyzed in this section.
Given a conserved spinor-tensor primary field in the $(l,\bar l)$ representation of the Lorentz group, with scaling dimension $\Delta$, we define
\be
(\partial\cdot f)_{\alpha_2\ldots \alpha_{l}}^{\dot{\beta}_2\ldots \dot{\beta}_{{\bar l}}}(x) \equiv (\epsilon \sigma^\mu)^{\alpha_1}_{\;\;\dot\beta_1}
\partial_\mu f_{\alpha_1\ldots \alpha_l}^{\dot{\beta}_1\ldots \dot{\beta}_{\bar l}}(x) = 0  \,.
\label{CurrCons}
\ee
Let us see how the 4D current conservation (\ref{CurrCons}) can be uplifted to 6D as a constraint on the field $F^{a_1\ldots a_l}_{b_1\ldots b_{\bar l}}$. This will allow us to work directly with the 6D invariants (\ref{eq:invar1})-(\ref{eq:invar4}), providing a great simplification. The analysis that follows is essentially a generalization to arbitrary conserved currents of the one made in ref.\cite{Costa:2011mg}, where only symmetric traceless
currents were considered. From eq.(\ref{fFrelation}), we get
\be
(\partial\cdot f)_{\alpha_2\ldots \alpha_{l}}^{\dot{\beta}_2\ldots \dot{\beta}_{{\bar l}}}(x) =  (X^+)^{\Delta-(l+\bar l)/2}  \partial_\mu \bigg( (e^\mu)_{a_1}^{\;b_1}
\mathbf{X}_{\alpha_2 a_2}\ldots  \mathbf{X}_{\alpha_{l} a_{l}}  \overline{\mathbf{X}}^{\dot\beta_2 b_2} 
\ldots \overline{\mathbf{X}}^{\dot\beta_{{\bar l}}b_{\bar l}} F^{a_1\ldots a_{l}}_{b_1\ldots b_{{\bar l}}}(X)\bigg) \,,
\ee
where  
\be
(e^\mu)_{a}^{\;b} \equiv - \mathbf{X}_{ a \alpha}  (\epsilon \sigma^\mu)^{\alpha}_{\;\;\dot\beta}   \overline{\mathbf{X}}^{\dot\beta b} = (M^{\mu +})_a^{\;b}\,,
\ee
in terms of the tensor
\be
M^{MN} = 2 \big( X^M \Sigma^N_{\;\; P} X^P - X^N \Sigma^M_{\;\; P} X^P \big)\,.
\ee
 Applying the derivative to each term gives
\be\begin{split}
(\partial\cdot f)_{\alpha_2\ldots \alpha_{l}}^{\dot{\beta}_2\ldots \dot{\beta}_{{\bar l}}} & =   (X^+)^{\Delta-(l+\bar l)/2} \mathbf{X}_{\alpha_3 a_3}\ldots  \mathbf{X}_{\alpha_{l} a_{l}}  \overline{\mathbf{X}}^{\dot\beta_3 b_3} 
\ldots \overline{\mathbf{X}}^{\dot\beta_{{\bar l}}b_{\bar l}}  \Bigg((\partial_\mu e^\mu)_{a_1}^{\;b_1} \mathbf{X}_{\alpha_2 a_2}  \overline{\mathbf{X}}^{\dot\beta_2 b_2}\!+ \frac{\partial X^M}{\partial x^\nu} (e^\nu)_{a_1}^{\;b_1}  \\
 & \bigg( (l-1)
\Big(\frac{\partial \mathbf{X}_{\alpha_{2}a_{2}}}{\partial X^M}\Big) \overline{\mathbf{X}}^{\dot\beta_2 b_2} +(\bar l-1)
\Big(\frac{\partial \overline{\mathbf{X}}^{\dot\beta_2 b_2}}{\partial X^M}\Big)  \mathbf{X}_{\alpha_{2}a_{2}} +  \mathbf{X}_{\alpha_{2}a_{2}} \overline{\mathbf{X}}^{\dot\beta_2 b_2}\frac{\partial}{\partial X^M}\bigg)  \Bigg) F^{a_1\ldots a_{l}}_{b_1\ldots b_{{\bar l}}}\,.
\label{ConsDetails}
\end{split}
\ee
After some algebraic manipulations, eq.(\ref{ConsDetails}) can be recast in the form
\be
(\partial\cdot f)_{\alpha_2\ldots \alpha_{l}}^{\dot{\beta}_2\ldots \dot{\beta}_{{\bar l}}}  =   (X^+)^{\Delta-(l+\bar l)/2+2} \mathbf{X}_{\alpha_2 a_2}\ldots  \mathbf{X}_{\alpha_{l} a_{l}}  \overline{\mathbf{X}}^{\dot\beta_2 b_2}  \ldots \overline{\mathbf{X}}^{\dot\beta_{{\bar l}}b_{\bar l}}  R^{a_2\ldots a_{l}}_{b_2\ldots b_{{\bar l}}}\,,
\ee
where
\be
R^{a_2\ldots a_{l}}_{b_2\ldots b_{{\bar l}}} = 2 \bigg( -\Big(X_M \Sigma^{MN} \frac{\partial}{\partial X^N}\Big)_{a_1}^{\;b_1} +\frac{1}{X^+} \Big(\Delta-\frac{l+\bar l}2-2\Big) X_M (\Sigma^{M+})_{a_1}^{\; b_1}\bigg)
F^{a_1\ldots a_{l}}_{b_1\ldots b_{{\bar l}}} \,.
\label{Rdef}
\ee
In writing eq.(\ref{Rdef}), we used the fact that $F$ is a homogeneous function of degree $\Delta + (l+\bar l)/2$ and the following two identities hold:
\be
\bigg( \Big(X_M \Sigma^{MN} \frac{\partial}{\partial X^N}\Big)_{a_1}^{\;b_1}  \mathbf{X}_{\alpha_{2}a_{2}} \bigg)   F^{a_1\ldots a_{l}}_{b_1\ldots b_{{\bar l}}} = 
\bigg( \Big(X_M \Sigma^{MN} \frac{\partial}{\partial X^N}\Big)_{a_1}^{\;b_1}  \overline{\mathbf{X}}^{\dot\beta_{2}b_{2}} \bigg)   F^{a_1\ldots a_{l}}_{b_1\ldots b_{{\bar l}}} = 0\,,
\ee
since $F$ is symmetric in its indices and satisfies eq.(\ref{notraceF}). 

Analogously to what found in ref.\cite{Costa:2011mg} for symmetric traceless operators, we see here what is special about operators that saturate the unitarity  bound (\ref{UnitarityBound}).
They are the only ones for which the 6D uplifted tensor $R$ is $SO(4,2)$ covariant. In our index-free notation, current conservation in 6D takes an extremely simple form:
\be
\partial\cdot f(x,s,\bar s) = 
(\partial\cdot  f(x))_{\alpha_2\ldots \alpha_l}^{\dot{\beta}_2\ldots \dot{\beta}_{\bar l}} s^{\alpha_2}
 \ldots s^{\alpha_l} \bar s_{\dot{\beta}_2}\ldots \bar s_{\dot{\beta}_{\bar l}}
= -2D\cdot F(X,S,\bar S) = 0 \,,
\label{ConserveddD}
\ee
where
\be\label{conservationD}
D = \Big(X_M \Sigma^{MN} \frac{\partial}{\partial X^N}\Big)_{a}^{\;b}  \frac{\partial}{\partial S_a} \frac{\partial}{\partial \bar S^b} \,.
\ee

\section{Example: Fermion - Fermion - Tensor Correlator}
\label{sec:example}

In this section we show some examples  on how to use the formalism presented in section \ref{sec:3pointFun}. 
In particular, we will determine all the three-point functions involving two fermion fields $\psi_\alpha$ and $\bar \chi^{\dot\beta}$:
$\langle \psi_\alpha(x_1) \bar \chi^{\dot\beta}(x_2) {\cal O}(x_3) \rangle$.\footnote{Three-point functions between two fermions and i) one scalar, ii) one vector, one rank two iii) symmetric or iv) antisymmetric tensor have already been considered in Appendix B of  ref.\cite{Dobrev:1973zg}, though some tensor structures in the correlators were missed in that paper.}  
According to eq.(\ref{eq:criterion2}), the only non-vanishing 3-point function occurs when ${\cal O}$ is in one of the 
following three  Lorentz representations: $(l,l)$, $(l+2,l)$ and $(l,l+2)$, with $l\geq 0$. 
We will determine the form of the correlators  in two cases: with non-conserved and conserved operator ${\cal O}$.

\subsection{Non-conserved Tensor}

Let us start by considering the $(l,l)$ representations.
According to eq. (\ref{eq:ff3pf}),  for $l=0$ there is only one possible structure to this correlator, proportional to $I_{21}$.
Using eqs.(\ref{f4dExp}) and (\ref{IijExp})  we immediately get
\be
\langle \psi_\alpha(x_1) \bar \chi^{\dot\beta}(x_2) \phi(x_3) \rangle = \lambda_{\psi\bar\chi{\cal O}_{0,0}} x_{12}^{-\Delta_{123}-1} x_{13}^{-\Delta_{132}} x_{23}^{-\Delta_{231}}
 (x_{21}\cdot \sigma\epsilon)_\alpha^{\;\;\dot\beta} \,,
\label{3pfscalar}
\ee
with $ \lambda_{\psi\bar\chi{\cal O}_{0,0}} $ a complex parameter. For $l\geq 1$, two independent structures are present,
\be\label{conservedten1}
\langle F_1 F_2 F_3 \rangle=\mathcal{K} \Big( \lambda_1  I_{21}J_{3,12}+ \lambda_2 I_{31} I_{23}\Big)J_{3,12}^{l-1}
\,, \ \ l\geq 1\,,
\ee
where $\lambda_{1,2}$ are two complex parameters and $\mathcal{K}$ is defined in eq.(\ref{eq:kinematicfactor1}).
Again using eqs.(\ref{f4dExp}),(\ref{IijExp}) and (\ref{JijExp}) we find
\be\begin{split}
\langle \psi_\alpha(x_1) \bar \chi^{\dot\beta}(x_2) t_{\alpha_1\ldots \alpha_l}^{\dot\beta_1\ldots\dot\beta_l}(x_3) \rangle    = &  \frac{ x_{12}^{-\Delta_{123}-l-1} x_{13}^{-\Delta_{132}+l} x_{23}^{-\Delta_{231}+l}}{(l!)^2}
\times \\
& \Big( \lambda^{(1)}_{\psi\bar\chi {\cal O}_{l,l}} (x_{12}\cdot \sigma\epsilon)_\alpha^{\;\dot\beta}  (Z_{3,12}\cdot\sigma\epsilon)_{\alpha_1}^{\;\dot\beta_1} \ldots  (Z_{3,12}\cdot\sigma\epsilon)_{\alpha_l}^{\;\dot\beta_l} +\frac{x_{12}^2}{2x_{13}^2x_{23}^2} \\
&  \lambda^{(2)}_ {\psi\bar\chi  {\cal O}_{l,l}}    (x_{13}\cdot \sigma\epsilon)_\alpha^{\;\dot\beta_1} (x_{23}\cdot \sigma\epsilon)_{\alpha_1}^{\;\dot\beta}  (Z_{3,12}\cdot\sigma\epsilon)_{\alpha_2}^{\;\dot\beta_2} \ldots  (Z_{3,12}\cdot\sigma\epsilon)_{\alpha_l}^{\;\dot\beta_l} \\ & +  {\rm perms.} \Big)\,.
\label{3pfsymmtensor}
\end{split}
\ee
In eq.(\ref{3pfsymmtensor}), $\lambda^{(1)}_{\psi\bar\chi  {\cal O}_{l,l}}$ and $\lambda^{(2)}_{\psi\bar\chi  {\cal O}_{l,l}}$ are proportional to $\lambda^{1}$ and $\lambda^{2}$ in eq.(\ref{conservedten1}) respectively with the same proportionality factor,  $Z_{3,12}^\mu$ is defined in eq.(\ref{Zkmu}) and perms. refer to the $(l!)^2-1$ terms obtained by permuting the $\alpha_i$ and $\dot\beta_i$ indices. When $\bar\chi$ is the complex conjugate of $\psi$, namely $\bar\chi^{\dot\beta} = \bar\psi^{\dot\beta} = (\psi_\beta)^\dagger$ and the symmetric traceless tensor  components are real, 
the OPE coefficients $ \lambda^{(1,2)}_{\psi\bar\chi t_l} $ are either purely real or purely imaginary, depending on $l$. 
When $x_{1,2,3}^\mu$ are space-like separated, causality implies that the operators commute between each other \cite{Rattazzi:2008pe}. Taking $\beta=\alpha$ and $\beta_i=\alpha_i$, we then have
\be
\langle \psi_\alpha(x_1) \bar \psi^{\dot\alpha}(x_2) t_{\alpha_1\ldots \alpha_l}^{\dot\alpha_1\ldots\dot\alpha_l}(x_3) \rangle^* = 
-\langle \psi_\alpha(x_2) \bar \psi^{\dot\alpha}(x_1) t_{\alpha_1\ldots \alpha_l}^{\dot\alpha_1\ldots\dot\alpha_l}(x_3) \rangle \,.
\ee
Since $Z_{3,12}= - Z_{3,21}$ we get 
\be
( \lambda^{(1)}_{\psi\bar\psi  {\cal O}_{l,l}})^*= (-1)^l  \lambda^{(1)}_{\psi\bar\psi  {\cal O}_{l,l}} \,, \ \ \ \  (\lambda^{(2)}_{\psi\bar\psi  {\cal O}_{l,l}})^* = (-1)^l  \lambda^{(2)}_{\psi\bar\psi  {\cal O}_{l,l}} \,.
\ee
Let us now consider the parity transformations of eq.(\ref{3pfsymmtensor}). 
Parity maps the three-point function  $\langle \psi_\alpha(x_1) \bar \chi^{\dot\beta}(x_2) t_{\alpha_1\ldots \alpha_l}^{\dot\beta_1\ldots\dot\beta_l}(x_3) \rangle $ to the 
complex conjugate three-point function $\langle \bar\psi^{\dot\alpha}(\tilde x_1) \chi_{\beta}(\tilde x_2) t^{\dot\alpha_1\ldots \dot\alpha_l}_{\beta_1\ldots\beta_l}(\tilde x_3) \rangle$.
When $\bar\chi=\bar\psi$, and $\alpha_i=\beta_i$, the three-point function is mapped to itself, provided the exchange $x_1\leftrightarrow x_2$ and $\alpha\leftrightarrow \beta$.  The two structures appearing in eq.(\ref{3pfsymmtensor})
have the same parity transformations. If we impose parity conservation in the CFT and we choose a negative intrinsic parity for the traceless symmetric tensor,  $\eta_{{\cal O}}=-1$,
then the three-point function must vanish: 
$\lambda^{(1)}_{\psi\bar\psi t_l}= \lambda^{(2)}_{\psi\bar\psi t_l}=0$. For $\eta_{{\cal O}}=1$, instead, parity invariance does not give any constraint.

Let us next consider the $(l+2,l)$ representations. According to eq. (\ref{eq:ff3pf}),  there is only one possible structure to this correlator, for any $l$:
\be\label{conservedAten1}
\langle F_1 F_2 F_3 \rangle=\mathcal{K}  \lambda I_{23}K_{2,13} J_{3,12}^{l} \,,
\ee
that gives rise to the 4D correlator
\be\begin{split}
\langle \psi_\alpha(x_1) \bar \chi^{\dot\beta}(x_2) t_{\alpha_1\ldots \alpha_{l+2}}^{\dot\beta_1\ldots\dot\beta_l}(x_3) \rangle    = &  \frac{ x_{12}^{-\Delta_{123}-l-1} x_{13}^{-\Delta_{132}+l} x_{23}^{-\Delta_{231}+l-2}}{(l!)(l+2)!} \lambda_{\psi\bar\chi {\cal O}_{l+2,l}}  \bigg(  (x_{23}\cdot \sigma\epsilon)_{\alpha_{l+1}}^{\;\dot\beta} 
\times \\
&\Big((x_{12}^2+x_{23}^2-x_{13}^2)\epsilon_{\alpha\alpha_{l+2}}+
4x_{12}^\mu x_{23}^\nu
(\sigma_{\mu\nu}\epsilon)_{\alpha\alpha_{l+2}} \Big) \times \\
&  (Z_{3,12}\cdot\sigma\epsilon)_{\alpha_1}^{\;\dot\beta_1} \ldots  (Z_{3,12}\cdot\sigma\epsilon)_{\alpha_l}^{\;\dot\beta_l}  +  {\rm perms.} \bigg)\,,
\label{3pfnosymmtensor}
\end{split}
\ee
where $\lambda_{\psi\bar\chi {\cal O}}$ is proportional to $\lambda$ in eq.(\ref{conservedAten1}).

A similar analysis applies to the complex conjugate $(l,l+2)$ representations. The only possible 6D structure is 
\be\label{conservedAten3}
\langle F_1 F_2 F_3 \rangle=\mathcal{K}  \lambda I_{31} \overline K_{1,23} J_{3,12}^l\,,
\ee
and gives
\be\begin{split}
\langle \psi_\alpha(x_1) \bar \chi^{\dot\beta}(x_2) t_{\alpha_1\ldots \alpha_{l}}^{\dot\beta_1\ldots\dot\beta_{l+2}}(x_3) \rangle    = &  \frac{ x_{12}^{-\Delta_{123}-l-1} x_{13}^{-\Delta_{132}+l-2} x_{23}^{-\Delta_{231}+l}}{(l!)(l+2)!} \lambda_{\psi\bar\chi {\cal O}_{l,l+2}}  \bigg(  (x_{31}\cdot \sigma\epsilon)_{\alpha}^{\;\dot\beta_{l+1}} 
\times \\
&\Big((x_{12}^2+x_{13}^2-x_{23}^2)\epsilon^{\dot\beta\dot\beta_{l+2}}+
4x_{21}^\mu x_{13}^\nu
(\bar\sigma_{\mu\nu}\epsilon)^{\dot\beta\dot\beta_{l+2}} \Big) \times \\
&  (Z_{3,12}\cdot\sigma\epsilon)_{\alpha_1}^{\;\dot\beta_1} \ldots  (Z_{3,12}\cdot\sigma\epsilon)_{\alpha_l}^{\;\dot\beta_l}  +  {\rm perms.} \bigg)\,.
\label{3pfnosymmtensor2}
\end{split}
\ee
If $\chi=\psi$, as expected, eq.(\ref{3pfnosymmtensor2}) is mapped to eq.(\ref{3pfnosymmtensor}) under parity transformation. In particular, in a parity invariant CFT, we should have the same number of $(l,l+2)$ and 
conjugate $(l+2,l)$ fields, with 
\be
 \lambda_{\psi\bar\psi {\cal O}_{l+2,l}} = \eta_{{\cal O}}  \lambda_{\psi\bar\psi {\cal O}_{l,l+2}}\,.
\ee

\subsection{Conserved Tensor} 

Let us start by considering  $(l,l)$ representations. The scaling dimension of ${\cal O}$ is now fixed to be $\Delta_3 = l+2$.
Taking the divergence (\ref{conservationD}) of eq.(\ref{conservedten1}) and using  eqs.(\ref{eq:gauge2}) and (\ref{eq:Xperm}) gives
\be\label{conservedten2}
 D_3 F=\frac{\Delta_1-\Delta_2}2\frac{2\lambda_2(l-1)(l+2) I_{31} I_{23}J_{3,12}^{l-2}+[\lambda_2+2l(l+1)\lambda_1] I_{21}J_{3,12}^{l-1}}{X_{12}^{\frac{\Delta_1 +\Delta_2-2l-1}{2}} X_{13}^{\frac{\Delta_1 -\Delta_2+2l+2}{2}} X_{23}^{\frac{-\Delta_1 +\Delta_2+2l+2}{2}}} = 0\,,
\ee
where the subscript 3 in $D$ indicates that derivatives are taken with respect to $X_3$, $S_3$ and $\bar S_3$. 
Eq.(\ref{conservedten2}) has the correct form for a Fermion-Fermion-spin $(l-1)$ symmetric tensor, as it should, and is automatically satisfied if $\Delta_1=\Delta_2$. For $\Delta_1\neq\Delta_2$ we have
\be
2\lambda_2(l-1)(l+2)=0\,, \ \ \ \ \ 
\lambda_1=-\frac{\lambda_2}{2l(l+1)} \,.
\label{ExpConstraint}
\ee
For $l=1$ we get one independent structure in eq.(\ref{conservedten1}) with $\lambda_2=-4\lambda_1$. For $l>1$ eq.(\ref{ExpConstraint}) admits only the trivial solution
\be
\langle \psi_\alpha(x_1) \bar \chi^{\dot\beta}(x_2){\cal O}_{\alpha_1 \ldots \alpha_l}^{\dot\beta_1\ldots \dot\beta_l}(x_3) \rangle=0\,, \ \ \ \\ l>1 \,, \ \ \Delta_1\neq \Delta_2\,.
\ee

Let us next consider the $(l+2,l)$ representations,  where ${\cal O}^{(l+2,l)}$ is a conserved tensor with $\Delta_3 = l+3$, $l>0$.
The divergence (\ref{conservationD}) of eq.(\ref{conservedAten1}) gives now 
\be\label{conservedAten2}
D_3 F=-\frac{\lambda}2(\Delta_2-\Delta_1+1)\frac{2l(l+3) I_{23}K_{2,31} J_{3,12}^{l-1}}{X_{12}^{\frac{\Delta_1 +\Delta_2-2l-3}{2}} X_{13}^{\frac{\Delta_1 -\Delta_2+2l+4}{2}} X_{23}^{\frac{-\Delta_1 +\Delta_2+2l+4}{2}}}=0 \,.
\ee
For $\Delta_2=\Delta_1-1$ eq.(\ref{conservedAten2}) is automatically satisfied. When $\Delta_2\neq \Delta_1-1$,
there are no non-trivial solutions of eq.(\ref{conservedAten2})  for $l>0$:
\be
\langle \psi_\alpha(x_1) \bar \chi^{\dot\beta}(x_2) {\cal O}_{\alpha_1 \ldots \alpha_{l+2}}^{\dot\beta_1\ldots \dot\beta_l}(x_3) \rangle=0\,, \ \ \ \ \l>0\,, \ \ \ \Delta_2\neq \Delta_1-1\,.
\ee
A similar result applies for conserved  ${\cal O}^{(l,l+2)}$ operators.

We have checked that the current conservation condition (\ref{ConserveddD}) reproduces various results found in the literature. In particular we have verified that the
correlator of  three energy-momentum tensors, once permutations and current conservations are imposed, contains three independent structures, as found in ref.\cite{Osborn:1993cr}.

\section{Consistency with Crossing Symmetry: Counting Four-Point Function Structures}
\label{sec:crossing}

We have seen how three-point functions of spinor-tensors in arbitrary representations of the Lorentz group can be computed.
The most subtle step of the procedure is the identification of the independent tensor structures entering three-point functions. For the particular case of traceless symmetric tensors, we reproduce
the results of ref.\cite{Costa:2011mg}. But only a subset of the building blocks we have found enter traceless symmetric tensors, so more checks are welcome.
In this section we use four-point functions to show how our three-point function counting passes the highly non-trivial consistency check of crossing symmetry.
Recall that, by using the OPE, the number of independent structures $N_4$ entering a generic four-point function
\begin{equation}
\langle {\cal O}_1(x_1) {\cal O}_2(x_2) {\cal O}_3(x_3) {\cal O}_4(x_4) \rangle
\label{4ptFun}
 \end{equation}
can be put in one to one correspondence with those of the three-point functions. For instance by taking the limit $x_1\rightarrow x_2$ and $x_3\rightarrow x_4$, eq.(\ref{4ptFun}) 
schematically boils down to the sum of the two-point functions 
\begin{equation}
\sum_r \sum_{i,j} C_{ {\cal O}_1 {\cal O}_2 {\cal O}_r}^iC_{ {\cal O}_3 {\cal O}_4 {\cal O}_{\bar r}}^j
 \langle {\cal O}_r^i(x_2) {\cal O}_{\bar r}^j(x_4)  \rangle\,.
 \label{4ptFunOPE}
  \end{equation}
In eq.(\ref{4ptFunOPE}), $r$ runs over all possible representations that can appear simultaneously in the two OPE's ($\bar r$ being the complex conjugate one), while
$i$ and $j$ denote, for a given representation $r$, the possible independent OPE coefficients, one for each independent tensor structure.
All kinematic factors and tensor structures have been omitted for simplicity.
Denoting the number of structures in the three-point function $ \langle {\cal O}_i {\cal O}_j {\cal O}_{r} \rangle$ by $N_{3r}^{ij}$, we conclude that 
\begin{equation}
N_4 = \sum_r N_{3r}^{12} N_{3\bar r}^{34} \,.
 \label{4ptFunOPENumber}
  \end{equation}
On the other hand, it is clear that the very same number $N_4$ should be obtained by pairing the four operators in any other way:
\begin{equation}
 \sum_r N_{3r}^{12} N_{3\bar r}^{34}  = \sum_r N_{3r}^{14} N_{3\bar r}^{23} = \sum_r N_{3r}^{13} N_{3\bar r}^{24}  \,.
 \label{Crossing}
  \end{equation}
We will refer to the three OPE pairing in eq.(\ref{Crossing})  as the $s$, $t$ or $u$-channel, respectively.
We have numerically verified the validity of eq.(\ref{Crossing}) for any four-point function involving arbitrary non-conserved fermionic or bosonic operators with $0\leq l_i,\bar l_i  \leq 6$,
$i=1,\ldots 4$. Finding a closed analytic form of $N_{3r}^{ij}$ in the most general case is a laborious task, so we focus here on the case in which 
the external operators are all symmetric traceless. The number of independent structures appearing in the three-point function of two symmetric traceless operators
$(l_1,l_1)$ $(l_2,l_2)$ and one arbitrary $(l_x,\bar{l}_x)$ tensor has been found in eq.(\ref{eq:counting}). 
Using that formula and summing over all the possible representations that can be exchanged, we can obtain in a closed analytical form the number of independent
structures for any four-point function involving arbitrary non-conserved symmetric traceless operators. 

For simplicity, let us consider four symmetric traceless operators with $l_3=l_1$, $l_4=l_2$, and $l_1\leq l_2$. In the $u$-channel, the representations that can appear in the OPE are of the form $(l_x,l_x+\delta)$,
with $|\delta|\leq 4 l_1$.  In the $s$ and $t$-channel,  they are of the form $(l_x,l_x+\delta)$, with $|\delta|\leq 2( l_1+l_2)$. Using eq.(\ref{eq:counting}) and summing over all the representations exchanged,
we get the number of four-point function structures. For $l_2\leq 2 l_1$ we get
\begin{equation}\begin{split}
N_4^{l_2\leq  2 l_1}(l_1,l_2) =& \frac{1}{630} \Bigg( 630-35l_1^7+1518 l_2+1232l_2^2+364l_2^3+35 l_2^4+7l_2^5 \\
& -7 l_2^6+l_2^7+7l_1^6 (34l_2-1)+l_1^5 (175+84l_2-672 l_2^2)+\\
& 35 l_1^4 (1-26l_2-12 l_2^2+28 l_2^3) +70l_1^3(1+20l_2+64l_2^2+40l_2^3 \\
& -4l_2^4) + 14l_1^2 (88+447 l_2+600 l_2^2 + 200l_2^3 - 30l_2^4 + 6 l_2^5) \\
& +14 l_1(120+398 l_2 + 384 l_2^2 + 100 l_2^3 - 5 l_2^4+ 6 l_2^5 - l_2^6)\Bigg)\,.
 \end{split}
 \label{4pointv1}
\end{equation}
 For $l_2\geq  2 l_1$ we get
 \begin{equation}\begin{split}
N_4^{l_2\geq   2 l_1}(l_1,l_2) =& \frac{1}{210} \Bigg(210+592l_1+448l_1^2-126l_1^3-175l_1^4+133l_1^5+147l_1^6+31l_1^7  \\
& -70 l_2 (l_1+1)^4(l_1(2+l_1) -7)+420 (1+l_1)^4 l_2^2 + 140 l_2^3 (1+l_1)^4
\Bigg)\,. \end{split}
 \label{4pointv2}
\end{equation}
Although it is not obvious from their expressions, eqs.(\ref{4pointv1}) and (\ref{4pointv2}) always give rise to positive integers numbers, as they should, and agree for $l_2=2l_1$.
In both cases, we get the same formula by counting structures either in the $s$-, $t$-, or $u$-channel. We believe this is a highly non-trivial check supporting the validity of our approach.
Eqs.(\ref{4pointv1}) and (\ref{4pointv2}) count the total number of parity even and parity odd structures.\footnote{As explained in subsection \ref{subsec:parity}, by parity even and odd we mean the structures that are respectively allowed or forbidden when we impose parity conservation to a correlator where the product of the 4 intrinsic parities equals one. In this case, the parity odd structures  in vector notation are those  involving one $\epsilon_{\mu\nu\rho\sigma}$ tensor.} 
For illustration we also report the individual number of parity even ($N_{4+}$) and parity odd ($N_{4-}$) 
structures when $l_2 = l_1=l$, i.e. four traceless symmetric operators with the same spin. We get
 \begin{equation}\begin{split}
N_{4+}(l) & = \frac{(l+1)(l+2)}{630} \Bigg(315+l \Big(957+l\big(1361+l(1127+151 l (l+4))\big)\Big)\Bigg) \,, \\
N_{4-}(l) & = \frac{(l+1)l}{630} \Bigg(339+l \Big(1789+l\big(2985+l(2335+151 l (l+6))\big)\Big)\Bigg) \,.
 \label{4pointv3}
 \end{split}
\end{equation}
We have in particular $N_{4+}(1) =43$, $N_{4+}(2) =594$, $N_{4+}(3) =4174$. Our formula for $N_{4+}(l)$ does not agree with eq.(4.68) of ref.\cite{Costa:2011mg} for $l\geq 2$. The number of structures which is found by using eq.(4.68) of that paper is slightly bigger than what found using $N_{4+}$ in eq.(\ref{4pointv3}) (there is agreement between the two formulas only for  $l=1$). 
The same kind of mismatch is found for four-point functions featuring traceless symmetric operators with different spins. 
We believe that ref.\cite{Costa:2011mg} might have missed some relation between invariants, resulting in an overcounting of structures in four space-time dimensions. 

It is important to stress that the number of invariants above refer to the generic case of four {\it different} non-conserved operators. 
For identical operators, the obvious permutation symmetries should be imposed, resulting in a reduced number of tensor structures.
For any given correlation function, the constraints arising from conserved operators are easily worked out
using the results of section \ref{sec:CurrentCons}, but we have not tried to get an analytical general formula in this case.

\section{Conclusions}

We have computed in this paper the most general three point function occurring in a 4D CFT between  bosonic and fermionic primary fields in 
arbitrary representations of the Lorentz group. We have used the 6D embedding formalism in twistor space with an index free notation, as introduced in ref.\cite{SimmonsDuffin:2012uy}, 
to efficiently recast the result in  terms of 6D SU(2,2) invariants.  The most important equation of the paper is the compact 6D formula (\ref{eq:ff3pf}), from which any 4D correlator can easily be extracted.
The constraints arising from conserved operators take a very simple form, see eqs.(\ref{ConserveddD}) and (\ref{conservationD}), and can be solved within our formalism.
Once the number of independent tensor structures in three-point functions are known,  one can  compute the number of independent structures of higher point functions
by taking the OPE limit in pairs of operators. As a highly non-trivial check of our results, we have shown that this number is independent of the way the operators are paired in the OPE, 
as it should be by crossing symmetry. 

As one of many applications of our results, we have reported the closed form expression  (\ref{eq:counting}) for the number of tensor structures
in three-point functions of two symmetric traceless and another arbitrary operator. Such result, in turn, allows to analytically determine  
the number of tensor structures of four-point functions of traceless symmetric tensors, see eqs.(\ref{4pointv1}), (\ref{4pointv2}) and (\ref{4pointv3}).

Understanding three-point functions is the first crucial step to extend the conformal bootstrap beyond scalar four-point functions.
The methods used in this paper should allow to determine the conformal blocks associated to fields in arbitrary Lorentz representations
entering in arbitrary four-point functions, in terms of a number of  ``seed" conformal blocks, analogously to the way 
the results of ref.\cite{Costa:2011mg} allow to compute conformal blocks of symmetric traceless tensors 
entering in four-point functions of symmetric traceless tensors in terms of the known conformal blocks for scalar four-point functions \cite{Costa:2011dw}. 
It would also be nice to extend to arbitrary bosonic and fermionic fields the conjectured agreement that was found between the number of tensor 
structures in $n$-point functions of symmetric traceless operators in D dimensional CFTs and the number of independent  terms in $n$-point scattering amplitudes of massive higher spin particles 
in flat D+1 dimensional Minkowski space \cite{Costa:2011mg}.  The embedding twistor formalism developed in this paper should be able to address this point for the D=4 case. 
We hope to come back to these further applications in a separate publication.

\vskip 10pt

{\it Note added:} During the final stages of this work, ref.\cite{Costa:2014rya} appeared, where tensors with mixed symmetry are studied. 
Ref.\cite{Costa:2014rya} considers CFTs in arbitrary dimensions, but focuses on bosonic, non-conserved, operators only.
When a comparison is possible, the number of tensor structures computed in the examples considered in ref.\cite{Costa:2014rya} agrees with our results.\footnote{We thank Tobias Hansen for some clarifications about the results presented in ref.\cite{Costa:2014rya}.}

\section*{Acknowledgments}

We thank Alejandro Castedo Echeverri and Hugh Osborn for useful discussions. 

\appendix

\section{Notation and Conventions}
\label{app:Notation}

We follow the conventions of Wess and Bagger \cite{WB} for the two-component spinor algebra in 4D.
Six dimensional vector indices are denoted by $M,N,\ldots$, with $M=\{\mu,+,-\}$; four dimensional vector indices are denoted by $\mu,\nu,\ldots$; 
four-dimensional spinor indices are denoted by  dotted and undotted Greek letters, $\alpha, \beta,\ldots$, $\dot{\alpha},\dot{\beta},\ldots$;
six-dimensional spinor (twistor) indices are denoted by  $a,b,\ldots$, with $a=\{\alpha,\dot\alpha\}$. We use capital and small letters for 6D and 4D tensors;
in particular, 6D and 4D coordinates are denoted as $X^M$ and $x^\mu$, where $x^\mu = X^\mu/X^+$.

The conformal group $SO(4,2)$ is locally isomorphic to $SU(2,2)$. The spinorial representations ${\bf 4}_\pm$ of $SO(4,2)$ are mapped to the fundamental and anti-fundamental representations
of $SU(2,2)$. Roughly speaking, $SO(4,2)$ spinor indices turn into $SU(2,2)$ twistor indices. We denote by $V_a$ and $\overline W^a\equiv W^{\dagger b}  \rho_b^a$, where $\rho$ is the $SU(2,2)$
metric, twistors  transforming in the fundamental and anti-fundamental of $SU(2,2)$, respectively:
\be
V\rightarrow U V\,, \ \ \ \ \overline W\rightarrow \overline W\, \overline{U}\,.
\label{VUtrans}
\ee
In eq.(\ref{VUtrans}), $U$ and $\overline{U}\equiv \rho\; U^{\dagger} \rho$ satisfy the condition $\overline{U} U = U\overline{U}  = 1$.

The non-vanishing components of the 6D metric $\eta_{MN}$ and its inverse $\eta^{MN}$  in light-cone coordinates are 
\be
\eta_{\mu\nu} = \eta^{\mu\nu} = {\rm diag} (-1,1,1,1)\,, \ \ \ \eta_{+-} = \eta_{-+} = \frac 12\,,  \ \ \ \eta^{+-} = \eta^{-+} = 2\,.
\ee
Six dimensional Gamma matrices $\Gamma^M$ are constructed by means of the 6D matrices $\Sigma^M$ and $\overline \Sigma^M$, analogues of $\sigma^\mu$ and $\bar\sigma^\mu$ in 4D:  
\begin{equation}
\Gamma^M=\left(\begin{array}{cc}
0 &  \Sigma^M\\
 \overline{\Sigma}^M  & 0 
 \end{array}\right)\,,
 \label{G6D}
\end{equation}
obeying the commutation relation
\begin{equation}
\lbrace \Gamma^M,\Gamma^N \rbrace=2\eta^{MN} \,.
\end{equation}
It is very useful to choose a basis for the $\Sigma$ and $\bar\Sigma$ matrices where they are antisymmetric.
This is explicitly given by
\begin{equation}\begin{split}\label{eq:defrho}
\Sigma^M_{ab} = &  \left\{
\left(\begin{array}{cc}
    0                 & \sigma^\mu_{\alpha\dot{\gamma}}\epsilon^{\dot{\beta}\dot{\gamma}} \\
   - \bar{\sigma}^{\mu\dot{\alpha}\gamma}\epsilon_{\beta\gamma} & 0    \end{array}\right),       
\left(\begin{array}{cc}
   0 & 0 \\
     0  &2 \epsilon^{\dot{\alpha}\dot{\beta}} 
      \end{array}\right),       
\left(\begin{array}{cc}
    -2 \epsilon_{\alpha\beta} & 0 \\
     0  & 0  
           \end{array}\right)     
\right\} \,,  \\
\overline{\Sigma}^{Mac} = & 
 \left\{
\left(\begin{array}{cc}
    0                 & -\epsilon^{\alpha\gamma}\sigma^\mu_{\gamma\dot{\beta}} \\
    \epsilon_{\dot{\alpha}\dot{\gamma}}\overline{\sigma}^{\mu\dot{\gamma}\beta} & 0         
    \end{array}\right),       
\left(\begin{array}{cc}
    -2\epsilon^{\alpha\beta} & 0 \\
     0  &0 
         \end{array}\right)       
\left(\begin{array}{cc}
   0 & 0 \\
     0  & 2 \epsilon_{\dot{\alpha}\dot{\beta}} \\
    \end{array}\right)    
\right\} \,,\end{split}
\end{equation}
where, in order, $M=\{\mu,+,-\}$ in eq.(\ref{eq:defrho}). 
The 6D spinor Lorentz generators are defined as
\be\begin{split}
\Sigma^{MN} \; = \; &\frac 14 (\Sigma^M\overline \Sigma^N- \Sigma^N\overline \Sigma^M)\,, \\
\overline\Sigma^{MN} \; = \; & \frac 14 (\overline\Sigma^M \Sigma^N- \overline\Sigma^N \Sigma^M)\,.
\end{split}
\ee
Useful relations among the $\Sigma^M$ and $\overline\Sigma^M$ matrices, used repeatedly in the paper, are the following:
\begin{equation}\begin{split}
\overline{\Sigma}^{Mab}=& -\frac{1}{2}\epsilon^{abcd}\Sigma^M_{cd}, \ \ \ \ \ 
\Sigma^{M}_{ab}= -\frac{1}{2}\epsilon_{abcd}\overline{\Sigma}^{Mcd}, \\
  \Sigma^M_{ab}\Sigma_{Mcd}= & \, 2\epsilon_{abcd},   \ \ \ \ \ \ \
  \overline{\Sigma}^{Mab}\overline{\Sigma}_M^{cd}=2\epsilon^{abcd},  \\
  \Sigma^M_{ab}\overline{\Sigma}_M^{cd}=& -2(\delta_a^c\delta_b^d-\delta_a^d\delta_b^c)\,,
\end{split}
\end{equation}
where $\epsilon_{1234}=\epsilon^{1234}=+1$.

The 6D null cone is defined by
\be
X^2 = X^M X^N \eta_{MN} = 0 \Longrightarrow X^- = - \frac{X_\mu X^\mu}{X^+}\,.
\ee
On the null cone we have
\begin{equation}\label{eq:XYproduct}
X_1\cdot X_2= X_1^M X_2^N \eta_{MN} = -\frac{1}{2}X_1^+ X_2^+ (x_1-x_2)^\mu (x_1-x_2)_\mu \,,
\end{equation}
where $x^\mu=X^\mu/X^+$ are the standard 4D coordinates. We define
\be 
x_{ij}^\mu \equiv x_i^\mu-x_j^\mu\,, \ \ \ \ x_{ij}^2 \equiv x_{ij}^\mu x_{\mu,ij}\,.
\ee

Twistor space-coordinates are defined as 
\be
\mathbf{X}_{ab} \equiv  X_M \Sigma^M_{ab} = - \mathbf{X}_{ba} \,, \ \ \ \overline{\mathbf{X}}^{ab}  \equiv X_M \overline\Sigma^{Mab}= - \overline{\mathbf{X}}^{ba}\,.
\label{TwistorCoord}
\ee
A very useful relation is
\begin{equation}
\mathbf{X} \overline{\mathbf{X}} =X_MX_N\Sigma^M\overline{\Sigma}^N=\frac{1}{2}X_MX_N(\Sigma^M\overline{\Sigma}^N+\Sigma^N\overline{\Sigma}^M)=X_MX^M=X^2,
\end{equation}
and similarly  $\overline{\mathbf{X}} \mathbf{X}  = X^2$. One also has
\begin{equation}\label{eq:Xperm}
\mathbf{X}_1 \overline{\mathbf{X}}_2 +\mathbf{X}_2 \overline{\mathbf{X}}_1=\overline{\mathbf{X}}_1 \mathbf{X}_2+\overline{\mathbf{X}}_2 \mathbf{X}_1=2 X_1\cdot X_2\,.
\end{equation}
In the basis defined by eq.(\ref{eq:defrho}), we have
\begin{equation}\label{eq:thetasplitting}
 \begin{cases}
   \mathbf{X} _{\alpha\gamma}=-X^+\epsilon_{\alpha\gamma} \\
 \mathbf{X} _{\alpha}^{\;\;\dot{\gamma}}=-X_\mu\sigma^\mu_{\alpha\dot{\beta}}\epsilon^{\dot{\beta}\dot{\gamma}} \\
  \mathbf{X} ^{\dot{\alpha}}_{\;\;\gamma}=X_\mu\overline{\sigma}^{\mu\dot{\alpha}\beta}\epsilon_{\beta\gamma} \\
   \mathbf{X} ^{\dot{\alpha}\dot{\gamma}}=X^-\epsilon^{\dot{\alpha}\dot{\gamma}}
    \end{cases}
\;\;\;\; 
  \begin{cases}
    \overline{\mathbf{X}}^{\alpha\gamma}=-X^-\epsilon^{\alpha\gamma} \\
   \overline{\mathbf{X}}^{\alpha}_{\;\;\dot{\gamma}}=-X_\mu\epsilon^{\alpha\beta}\sigma^\mu_{\beta\dot{\gamma}} \\
   \overline{\mathbf{X}}_{\dot{\alpha}}^{\;\;\gamma}=X_\mu\epsilon_{\dot{\alpha}\dot{\beta}}\overline{\sigma}^{\mu\dot{\beta}\gamma} \\
   \overline{\mathbf{X}}_{\dot{\alpha}\dot{\gamma}}=X^+\epsilon_{\dot{\alpha}\dot{\gamma}}
 \end{cases}
\end{equation}
The 4D spinors are embedded as follows in the 6D chiral spinors (twistors):
\begin{equation}
    \Psi_{a}= \left(\begin{array}{c}
    \psi_{\alpha}     \\
   \bar \chi^{\dot{\alpha}}     
 \end{array}\right)\,, \ \ \ \ 
   \bar \Phi^{a}= \left(\begin{array}{c}
    \phi^{\alpha}     \\
    \bar\xi_{\dot{\alpha}}     
     \end{array}\right)\,.
 \end{equation}
In order to avoid a proliferation of spinor indices, we define
\be 
(\sigma^\mu\epsilon)_{\alpha}^{\dot\gamma}\equiv\sigma^\mu_{\alpha\dot\beta} \epsilon^{\dot\beta\dot\gamma}\,.
\label{sigmaepsilon}
\ee
Notice that in writing eq.(\ref{sigmaepsilon}) we have used the usual convention of matrix multiplication.
A similar comment applies for other similar expressions involving $\bar\sigma^\mu$, $\sigma^{\mu\nu}$ and $\bar\sigma^{\mu\nu}$.

\section{Spinor and Vector Notation for Tensor Fields}
\label{app:LorentzRep}

We usually write bosonic fields transforming in the lowest representations of the Lorentz group in vector notation: $A_\mu$, $T_{\mu\nu}$, etc.
With the notable exception of symmetric traceless tensors of the form $T_{(\mu_1 \ldots \mu_l)}$, the vector notation becomes awkward for higher spin.
On the contrary, by using the isomorphism between $SO(3,1)$ and $SL(2,C)$, a generic irreducible representation of the Lorentz group is 
defined by two integers $(l,\bar l)$. The matrix $\sigma^\mu$ provides the link between the vector and spinor representations of fields.
Given a reducible bosonic tensor field $t_{\mu_1\ldots \mu_n}$ or fermionic spinor-tensor fields $\psi_{\alpha,\mu_1\ldots \mu_n}$, $\bar\psi^{\dot\alpha}_{\mu_1\ldots \mu_n}$, we have 
\be\begin{split}
(\sigma^{\mu_1}\epsilon)^{\;\dot\beta_1}_{\alpha_1} \ldots (\sigma^{\mu_n}\epsilon)^{\;\dot\beta_n}_{\alpha_n} t_{\mu_1\ldots \mu_n} =  \, & \sum_{l,\bar l}^n t_{\alpha_1\ldots \alpha_l}^{\dot\beta_1\ldots \beta_{\bar l}}
\epsilon_{\alpha_{l+1}\alpha_{l+2}} \ldots \epsilon_{\alpha_{n-1}\alpha_{n}} \epsilon^{\dot\beta_{\bar l+1}\dot\beta_{\bar l+2}} \ldots \epsilon^{\dot\beta_{n-1}\dot\beta_{n}} \,, \\
(\sigma^{\mu_1}\epsilon)^{\;\dot\beta_1}_{\alpha_1} \ldots (\sigma^{\mu_n}\epsilon)^{\;\dot\beta_n}_{\alpha_n} \psi_{\gamma\mu_1\ldots \mu_n} =\, & \sum_{l,\bar l}^n \psi_{\gamma\alpha_1\ldots \alpha_l}^{\dot\beta_1\ldots \beta_{\bar l}}
\epsilon_{\alpha_{l+1}\alpha_{l+2}} \ldots \epsilon_{\alpha_{n-1}\alpha_{n}} \epsilon^{\dot\beta_{\bar l+1}\dot\beta_{\bar l+2}} \ldots \epsilon^{\dot\beta_{n-1}\dot\beta_{n}} \,,\\
(\sigma^{\mu_1}\epsilon)^{\;\dot\beta_1}_{\alpha_1} \ldots (\sigma^{\mu_n}\epsilon)^{\;\dot\beta_n}_{\alpha_n} \bar\psi^{\dot\gamma}_{\mu_1\ldots \mu_n} = \, & \sum_{l,\bar l}^n \bar\psi_{\alpha_1\ldots \alpha_l}^{\dot\gamma\dot\beta_1\ldots \beta_{\bar l}}
\epsilon_{\alpha_{l+1}\alpha_{l+2}} \ldots \epsilon_{\alpha_{n-1}\alpha_{n}} \epsilon^{\dot\beta_{\bar l+1}\dot\beta_{\bar l+2}} \ldots \epsilon^{\dot\beta_{n-1}\dot\beta{n}} 
\,,\end{split}
\label{TSpVect}
\ee
where 
the sum over $l$, $\bar l$ runs over even or odd integers, for even or odd $n$, respectively. 
Taking symmetric and antisymmetric combinations in the undotted and dotted indices of the r.h.s. of eq.(\ref{TSpVect}) allows us to find the explicit relations between the different field components in vector and spinor notations. Inverse relations are obtained by multiplying eq.(\ref{TSpVect}) by powers of $(\epsilon\sigma^\mu)$:
\be\begin{split}
t_{\mu_1\ldots \mu_n} = 2^{-n}\, &  \sum_{l,\bar l}^n (\epsilon\sigma_{\mu_1})^{\alpha_1}_{\;\dot\beta_1}\ldots  (\epsilon\sigma_{\mu_n})^{\alpha_n}_{\;\dot\beta_n}t_{\alpha_1\ldots \alpha_l}^{\dot\beta_1\ldots \dot\beta_{\bar l}}
\epsilon_{\alpha_{l+1}\alpha_{l+2}} \ldots \epsilon_{\alpha_{n-1}\alpha_{n}} \epsilon^{\dot\beta_{\bar l+1}\dot\beta_{\bar l+2}} \ldots \epsilon^{\dot\beta_{n-1}\dot\beta_{n}}  \,, \\
\psi_{\gamma\mu_1\ldots \mu_n} = 2^{-n}\, &  \sum_{l,\bar l}^n (\epsilon\sigma_{\mu_1})^{\alpha_1}_{\;\dot\beta_1}\ldots  (\epsilon\sigma_{\mu_n})^{\alpha_n}_{\;\dot\beta_n}\psi_{\gamma\alpha_1\ldots \alpha_l}^{\dot\beta_1\ldots \dot\beta_{\bar l}}
\epsilon_{\alpha_{l+1}\alpha_{l+2}} \ldots \epsilon_{\alpha_{n-1}\alpha_{n}} \epsilon^{\dot\beta_{\bar l+1}\dot\beta_{\bar l+2}} \ldots \epsilon^{\dot\beta_{n-1}\dot\beta_{n}}  \,, \\
\bar\psi^{\dot\gamma}_{\mu_1\ldots \mu_n} = 2^{-n}\, &  \sum_{l,\bar l}^n (\epsilon\sigma_{\mu_1})^{\alpha_1}_{\;\dot\beta_1}\ldots  (\epsilon\sigma_{\mu_n})^{\alpha_n}_{\;\dot\beta_n}\bar\psi_{\alpha_1\ldots \alpha_l}^{\dot\gamma\dot\beta_1\ldots \dot\beta_{\bar l}}
\epsilon_{\alpha_{l+1}\alpha_{l+2}} \ldots \epsilon_{\alpha_{n-1}\alpha_{n}} \epsilon^{\dot\beta_{\bar l+1}\dot\beta_{\bar l+2}} \ldots \epsilon^{\dot\beta_{n-1}\dot\beta_{n}}  \,.
\label{TSpVectInv}\end{split}
\ee
It may be useful to work out in detail the case for, say,  a bosonic rank-two tensor $t_{\mu\nu}$. We have
\be
(\sigma^\mu\epsilon)^{\;\dot\beta_1}_{\alpha_1} (\sigma^\nu\epsilon)^{\;\dot\beta_2}_{\alpha_2} t_{\mu\nu} =  t \epsilon_{\alpha_1\alpha_2} \epsilon^{\dot\beta_1\dot\beta_2}+  t_{\alpha_1\alpha_2}\epsilon^{\dot\beta_1\dot\beta_2}
+  t^{\dot\beta_1\dot\beta_2}\epsilon_{\alpha_1\alpha_2}+ t^{\dot\beta_1\dot\beta_2}_{\alpha_1\alpha_2} \,,
\label{4DLor}
\ee
which corresponds to the decomposition $(0,0)\oplus (1,0)\oplus (0,1) \oplus (1,1)$, scalar, self-dual antisymmetric tensor, anti self-dual antisymmetric tensor, symmetric tensor.  From eq.(\ref{4DLor}) we get
\be\begin{split}
t \; = \; & \frac 12 \eta^{\mu\nu}t_{\mu\nu} \,, \\
t_{\alpha_1\alpha_2}  \; = \;  \, &  t_{\mu\nu} (\sigma^{\mu\nu} \epsilon)_{\alpha_1\alpha_2} \,, \\
t^{\dot\beta_1\dot\beta_2}  \; = \;  \, & t_{\mu\nu} (\epsilon\bar\sigma^{\mu\nu})^{\dot\beta_1\dot\beta_2} \,, \\
t_{\alpha_1\alpha_2}^{\dot\beta_1\dot\beta_2}  \; = \;  & \frac 14 t_{\mu\nu} \Big( (\sigma^\mu \epsilon)_{\alpha_1}^{\;\dot\beta_1} (\sigma^\nu \epsilon)_{\alpha_2}^{\; \dot\beta_2}+ (\sigma^\mu \epsilon)_{\alpha_2}^{\; \dot\beta_1} (\sigma^\nu \epsilon)_{\alpha_1}^{\; \dot\beta_2} + (\mu \leftrightarrow \nu) \Big) \,.
 \end{split}
\label{SpinVectRel}
\ee
Notice that in the last relation in eq.(\ref{SpinVectRel}) the trace part of $t_{\mu \nu}$ automatically gives a vanishing contribution. 
We get the inverse relations by means of eq.(\ref{TSpVectInv}). Decomposing $t_{\mu\nu}=\eta_{\mu\nu}t/2+t_{[{\mu\nu}]}+t_{(\mu\nu)}$,
where $t_{(\mu\nu)} = 1/2 (t_{\mu\nu}+t_{\nu\mu})-\eta_{\mu\nu} t/2$ and  $t_{[\mu\nu]} = 1/2 (t_{\mu\nu}-t_{\nu\mu})$, one has
\be\begin{split}
t_{[\mu\nu]} = \;&  \frac 12(\epsilon\sigma_{\mu\nu})^{\alpha_1\alpha_2} t_{\alpha_1\alpha_2} +\frac 12(\bar\sigma_{\mu\nu}\epsilon)_{\dot\beta_1\dot\beta_1} t^{\dot\beta_1\dot\beta_2}\,, \\
t_{(\mu\nu)} = \; &(\epsilon\sigma_\mu)_{\;\dot\beta_1}^{\alpha_1}(\epsilon\sigma_\nu)_{\;\dot\beta_2}^{\alpha_2} t_{\alpha_1\alpha_2}^{\dot\beta_1\dot\beta_2} \,.
\end{split}
\ee
For arbitrary symmetric traceless fields $t_{(\mu_1\ldots \mu_l)}$,  in particular, we have
\be\begin{split}
t^{\dot\beta_1\ldots \dot\beta_l}_{\alpha_1\ldots \alpha_l} = & \frac{1}{l!} t_{(\mu_1\ldots \mu_l)} \Big( (\sigma^{\mu_1} \epsilon)_{\alpha_1}^{\;\dot\beta_1} \ldots  (\sigma^{\mu_l} \epsilon)_{\alpha_l}^{\;\dot\beta_l} + {\rm perms.}\Big)\,, \\
 t_{(\mu_1\ldots \mu_l)} = & (\epsilon\sigma_{\mu_1})_{\;\dot\beta_1}^{\alpha_1}\ldots (\epsilon\sigma_{\mu_l})_{\;\dot\beta_l}^{\alpha_l} t^{\dot\beta_1\ldots \dot\beta_l}_{\alpha_1\ldots \alpha_l} \,.
\label{ChangebasisTrace}
\end{split}\ee

\end{document}